\documentclass[lettersize,journal]{IEEEtran}
\IEEEoverridecommandlockouts
\usepackage{cite}
\usepackage{color}
\usepackage{amsthm}
\usepackage{makecell}
\usepackage{amsmath,amssymb,amsfonts}
\usepackage{bbm}
\usepackage{dsfont}
\usepackage{algorithmic}
\usepackage{graphicx}
\usepackage{textcomp}
\usepackage{xcolor}
\usepackage{verbatim}
\usepackage{amsthm,amsmath,amssymb}
\usepackage{steinmetz}
\usepackage{textcomp}
\usepackage{mathrsfs}
\usepackage{amsmath,bm}
\usepackage{algorithm} 
\usepackage{mathtools}
\usepackage{epsf}
\usepackage{cleveref}
\usepackage{verbatim}
\usepackage{subfigure}
\usepackage{tikz}
\usepackage[justification=centering]{caption} 

\newtheorem{thm}{Theorem}
\newtheorem{lem}{Lemma}
\newtheorem{prop}{Proposition}
\allowdisplaybreaks
\def\BibTeX{{\rm B\kern-.05em{\sc i\kern-.025em b}\kern-.08em
    T\kern-.1667em\lower.7ex\hbox{E}\kern-.125emX}}
\usepackage{xcolor}
\usepackage{xpatch}
\makeatletter
\begin{document}

\title{
Modulation Design and Optimization for RIS-Assisted Symbiotic Radios
}
\author{
\IEEEauthorblockN{Hu~Zhou,~Bowen~Cai,~Qianqian~Zhang,~Ruizhe~Long,~Yiyang~Pei,\\and~Ying-Chang Liang,~\IEEEmembership{Fellow,~IEEE}}
\thanks{This work has been submitted to the IEEE for possible publication. Copyright may be transferred without notice, after which this version may no longer be accessible.}
\thanks{
Part of this work has been presented in IEEE Globecom 2022 \cite{zhou2021}.
H.~Zhou, Q.~Zhang, R.~Long, and Y.-C.~Liang are with the University of Electronic Science and Technology of China (UESTC), Chengdu 611731, China (e-mail: {huzhou@std.uestc.edu.cn;~qqzhang\_kite@163.com;~ruizhelong@gmail.com;~liangyc@ieee.org}). B.~Cai is with the China Telecom Research Institute, Beijing 102209, China (email: {caibw@chinatelecom.cn}). Y. Pei is with the Singapore Institute of Technology, 138683, Singapore (e-mail: yiyang.pei@singaporetech.edu.sg).}
}
\maketitle
\begin{abstract}
In reconfigurable intelligent surface (RIS)-assisted symbiotic radio (SR), the RIS acts as a secondary transmitter by modulating its information bits over the incident primary signal and simultaneously assists the primary transmission, then a cooperative receiver is used to jointly decode the primary and secondary signals. Most existing works of SR focus on using RIS to enhance the reflecting link while ignoring the ambiguity problem for the joint detection caused by the multiplication relationship of the primary and secondary signals.
Particularly, in case of a blocked direct link, joint detection will suffer from severe performance loss due to the ambiguity, when using the conventional on-off keying and binary phase shift keying modulation schemes for RIS. To address this issue, we propose a novel modulation scheme for RIS-assisted SR that divides the phase-shift matrix into two components: the symbol-invariant and symbol-varying components, which are used to assist the primary transmission and carry the secondary signal, respectively. To design these two components, we focus on the detection of the composite signal formed by the primary and secondary signals, through which a problem of minimizing the bit error rate (BER) of the composite signal is formulated to improve both the BER performance of the primary and secondary ones. By solving the problem, we derive the closed-form solution of the optimal symbol-invariant and symbol-varying components, which is related to the channel strength ratio of the direct link to the reflecting link. Moreover, theoretical BER performance is analyzed.
Finally, simulation results show the superiority of the proposed modulation scheme over its conventional counterpart.
\end{abstract}
\newcommand\degree{^\circ}
\section{Introduction}
The sixth generation (6G) mobile communications are envisioned to provide the ultimate and immersive experience for humanity through the hyper-connection of everything~\cite{SamsungPaper}. 
With the rapid emergence of wireless applications (e.g., eXtended reality),
more and more devices will be connected to networks, which requires plenty of spectrum and energy resources~\cite{you2021towards,latva2020key}. 
Recently, symbiotic radio (SR)~\cite{long2019symbiotic,liang2020symbiotic} has emerged as a promising candidate solution due to its mutualistic spectrum- and energy-sharing features\cite{nawaz2021non,janjua2023survey}. In SR, there co-exist two types of transmissions. Specifically, the primary transmission is achieved by using active radio technology with the allocated spectrum, e.g., cellular communication. 
By sharing the spectrum and energy with the primary one, the secondary transmission is realized by using the passive radio technology, e.g., backscatter communication~\cite{griffin2008gains,boyer2013backscatter,kimionis2014increased,liu2013ambient,van2018ambient,wang2024multi}, in which the secondary transmitter (STx) sends the secondary signal by modulating its information bits over the primary signal emitted by the primary transmitter (PTx). Due to the nature of backscatter modulation~\cite{kimionis2014increased}, the backscattered signal out of the STx is the multiplication of primary and secondary signals.
To decode the two signals, joint decoding is adopted at the cooperative receiver (C-Rx). By doing so, the secondary transmission can be achieved without dedicated spectrum and power-hungry RF chains, and in return, the secondary transmission can provide a multi-path for the primary transmission, thereby yielding mutualistic benefits for both. Due to its spectrum and energy-efficient features, SR has attracted increasing attention in recent years. 

In the literature, the mutualistic mechanism of SR is analyzed in~\cite{zhang2022mutualistic}, and it gives the condition under which the primary transmission could reap the benefits brought by the secondary transmission. The cell-free SR system is studied in~\cite{dai2022rate} which proposes a practical channel estimation method and optimizes the PTx beamforming. Also, the authors of \cite{xu2023MIMO} focus on the achievable rate analysis, while the authors of \cite{chen2020stochastic,long2019symbiotic,chu2020resource} study the resource allocation for SR. 
Note that in the above works, the classical monostatic/bistatic/ambient backscatter modulation schemes offer valuable insights for efficiently delivering the secondary signal.
These schemes include but are not limited to on-off keying (OOK)~\cite{kimionis2012bistatic,kimionis2014increased,wang2016ambient}, phase-shift keying (PSK)~\cite{griffin2008gains,bharadia2015backfi,darsena2017modeling,wang2012efficient}, frequency-shift keying (FSK)~\cite{kimionis2013bistatic,fasarakis2015coherent,wang2017fm,vougioukas2018coherent}, and quadrature amplitude modulation (QAM)~\cite{boyer2013backscatter,thomas2010qam,thomas2012quadrature}. For example, based on the PSK scheme, the achievable rate region of the SR is characterized in~\cite{liu2018backscatter}, where the corresponding detection scheme and performance analysis are also conducted. Different from backscatter communication which focuses on decoding the secondary signal, the C-Rx in SR jointly decodes the primary and secondary signals.
However, joint decoding in SR suffers from the ambiguity problem due to the multiplication of the primary and secondary signals. Particularly, when the direct link is blocked, the C-Rx would fail to perform joint decoding due to ambiguity. Moreover, due to the double fading effect (i.e., the double path losses from PTx to STx and from STx to C-Rx), the reflecting link is weak, which thus limits the performance of secondary transmission and the enhancement to the primary transmission.

Fortunately, reconfigurable intelligent surface (RIS)~\cite{wu2019towards} can be
introduced into SR as an STx to enhance the reflecting link by passive beamforming~\cite{lei2021reconfigurable,liang2022backscatter,zhang2024channel}. In RIS-assisted SR, RIS needs to play the role of STx to simultaneously transmit the secondary signal and assist the primary transmission~\cite{zhou2023assistance}.
Currently, most of the existing works on RIS-assisted SR focus on the joint active and passive beamforming design to achieve a variety of goals, such as transmit power minimization~\cite{zhang2021reconfigurable,zhou2022cooperative,zhang2023interference}, achievable rate maximization~\cite{wang2021intelligent,hu2020reconfigurable}, and BER minimization~\cite{hua2021novel,hua2021uav}. Whereas only a few works study the modulation design of RIS-assisted SR~\cite{zhang2021reconfigurable,lin2021reconfigurable,guo2020reflecting,Wu2021Reconfigurable,li2022reconfigurable,yan2020passive}. To be specific, in~\cite{hua2021uav,yan2020passive}, the OOK modulation scheme is investigated to send the secondary signal by adjusting the on/off states of the RIS reflecting elements. To avoid the power loss of the off elements in OOK modulation, RIS-aided quadrature reflection
modulation scheme is proposed in~\cite{lin2021reconfigurable} by partitioning the RIS into two orthogonal subsets, where the secondary signal is embedded into different RIS partitions. Besides, in~\cite{zhang2021reconfigurable,zhou2022cooperative}, the binary phase shift keying (BPSK) modulation scheme is studied by adjusting two types of phase shifts with a phase difference of $\pi$. Moreover, given a reflecting pattern candidate set, a general reflection modulation framework is proposed in~\cite{guo2020reflecting}, by jointly designing the signal mapping, shaping, and reflecting phase shifts. In~\cite{Wu2021Reconfigurable}, a symbiotic spatial modulation framework is proposed to transmit the secondary signal including one coherent modulation scheme and two non-coherent modulation schemes, under which the theoretical BER performance is analyzed. In~\cite{li2022reconfigurable}, the number modulation scheme is developed by dividing the RIS into two in-phase 
and quadrature subsets, where the number of reflecting elements in one subset is utilized to deliver the secondary signal.

Although the introduction of RIS into SR brings many benefits and opportunities, such as compensating for the double fading effect and enabling diverse modulation schemes, the ambiguity problem of SR has not been well addressed, which is unveiled in most existing studies. Particularly, when RIS is introduced, the impact of the ambiguity problem will be more pronounced since the reflecting link enhanced by RIS can even be stronger than the direct link with the passive beamforming gain. In this case, the BER performance of joint decoding in SR will be dominated by the ambiguity associated with the stronger reflecting link, which leads to severe performance loss and therefore motivates our work.

To address the ambiguity problem, in this paper, we propose a novel modulation scheme for RIS-assisted SR. The proposed scheme divides the RIS phase-shift matrix into two components: the symbol-invariant component used to assist the primary transmission and the symbol-varying component used to carry the secondary signal. Under the modulus constraints of phase shifts, there exists a fundamental tradeoff between the BER performance of the primary and secondary transmissions in terms of the design of the symbol-invariant and symbol-varying components. To study the tradeoff, this paper aims to optimize these two components to improve both the BER performance of the primary and secondary transmissions. In addition, the reflecting modulation~\cite{guo2020reflecting}, spatial modulation~\cite{Wu2021Reconfigurable}, and number modulation~\cite{li2022reconfigurable} use different reflection patterns or antenna indices to convey RIS information, which can also tackle the ambiguity problem of SR and provide insightful ideas. Different from them, our proposed modulation design can not only address the ambiguity problem by regarding the symbol-invariant component as a virtual direct link, but also help enhance the primary transmission when the direct link is weak. Our main contributions are summarized as follows:
\begin{itemize}
	\item We propose a novel modulation scheme for RIS-assisted SR, which divides the RIS phase-shift matrix into two components: the symbol-invariant component used to assist the primary transmission and the symbol-varying component used to carry the secondary signal. 
	\item To jointly decode signals at the C-Rx, instead of focusing on the individual primary and secondary signals~\cite{zhang2022mutualistic}, we particularly focus on the detection of the composite signal formed by the primary and secondary ones. Then, a two-step joint detector is developed by first decoding the composite signal, followed by the bit mapping from the composite signal to the primary and secondary signals. 
	\item To optimize the symbol-invariant and symbol-varying components, we aim to minimize the BER of the composite signal such that both the BER performance of the primary and secondary transmissions can be improved.
    Although it is challenging to solve the formulated problem due to its non-convexity, we derive the closed-form solution by using the geometrical analysis. 
       \item  It is shown that the optimal design of the symbol-invariant and symbol-varying components is related to the channel strength ratio of the direct link to the reflecting link, which determines the energy allocated to the symbol-invariant and symbol-varying components.
       Moreover, performance analysis is conducted to draw insights into the interrelationships between the primary and secondary transmissions.
	\item Extensive simulation results are presented to verify the superiority of the proposed modulation scheme, and it is shown that the proposed modulation scheme could strike a balance between the BER performance of the primary and secondary transmissions.
\end{itemize}


The rest of this paper is organized as follows. 
Section \ref{sec-system-model} introduces the system model. Section \ref{sec: proposed-modulation-scheme} presents the proposed modulation scheme and the receiver design.
Section \ref{sec-modulation-design} presents design criteria and problem formulation. Section \ref{sec: geometrical-analysis} utilizes the geometrical analysis to solve the formulated problem. Section \ref{sec-performance-analysis} provides the theoretical BER analysis. Section \ref{sec-simulation} presents the simulation results to evaluate our proposed scheme. Finally, Section \ref{sec-con} concludes the paper.

The main notations are listed as follows. The scalar is denoted by the lowercase letters.  $\mathbb{C}^{x \times y} $ denotes the space of $ x \times y $ complex-valued matrices.
$\mathcal{CN}(\mu,\sigma^{2})$ denotes complex Gaussian distribution with mean $\mu$ and variance $\sigma^{2}$. 
$\jmath$ denotes the imaginary unit. $\angle x$, $x^{H}$, $\Re\{x\}$, and $\Im\{x\}$ denote the phase, the conjugate transpose, the real and the imaginary parts of the complex number $x$, respectively. $\mathrm{root}(\cdot)$ denotes the roots of the equation. $\mathrm{erf}(u)\triangleq\frac{2}{\sqrt{\pi}}\int_{0}^{u}e^{-t^{2}}dt$ is the Gaussian error function. $\bm{0}_{K\times K}$ denotes a $K \times K$ matrix whose all elements are zero. $\bm{I}_{K\times K}$ denotes a $K\times K$ identity matrix. $[\bm{\Phi}]_{i,j}$ denotes the $(i,j)$-th element of the matrix $\bm{\Phi}$.
\begin{figure}  
	\centering  
	\setlength{\abovecaptionskip}{0.cm}
	\includegraphics[width=3in]{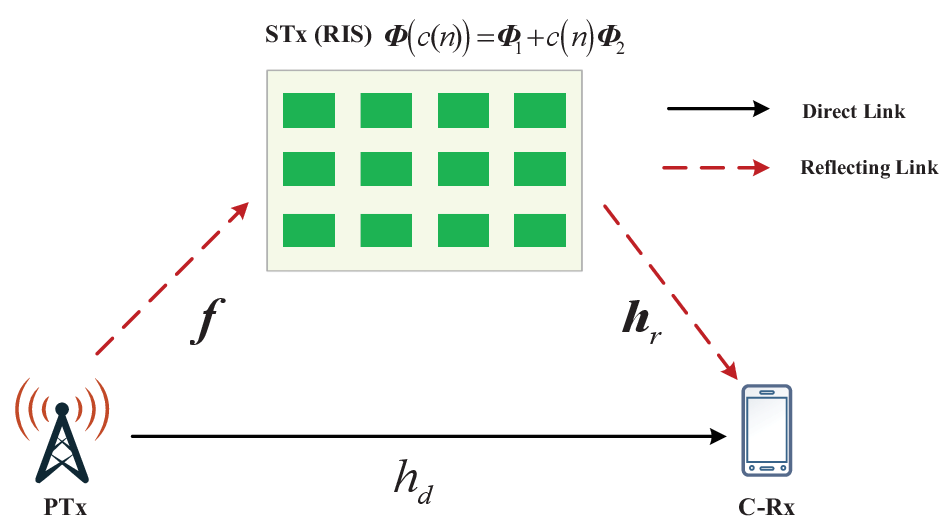}  
	\caption{System model of RIS-assisted SR.}  
	\label{fig:system-model}  
\end{figure}
\vspace{-0.4cm}
\section{System Model and Preliminary}
\label{sec-system-model}
As shown in Fig. \ref{fig:system-model}, we consider an RIS-assisted SR system, which consists of an active single-antenna PTx\footnote{Our proposed modulation scheme can be readily extended to the scenario with multi-antenna PTx, where the transmit beamforming and modulation design need to be jointly optimized to improve the performance of SR.}, an RIS with $K$ reflecting elements, and a single-antenna C-Rx. 
In such a system, RIS plays two roles. On one hand, it acts as a helper to assist the primary transmission by passive beamforming. On the other hand, RIS is connected with some sensors for collecting environmental data, and it acts as an STx to transmit the secondary signal (i.e., sensor information) by modulating it over the incident primary signal. Here, the task of C-Rx is to jointly decode both the primary and secondary signals.
In the following, we will first introduce the signal model, and then review the preliminary of the conventional modulation schemes.
\vspace{-0.2cm}
\subsection{Signal Model}
The quasi-static flat-fading model is considered. 
The channel response from the PTx to the C-Rx, from the PTx to the RIS, and from the RIS to the C-Rx are denoted by $h_{d}\in\mathbb{C}$, $\bm{f}\in\mathbb{C}^{K\times 1}$, and $\bm{h}_{r}\in\mathbb{C}^{K\times 1}$, respectively. For simplicity, we refer to the PTx--C-Rx link and the PTx--RIS--C-Rx link as the direct link and the reflecting link, respectively.
Denote by $\mathcal{A}_{s}$ and $\mathcal{A}_{c}$ the normalized constellation sets of the primary and secondary signals, respectively. 
Let $p$ denote the transmit power, $s(n)\in \mathcal{A}_{s}$ denote the $n$-th primary symbol, and $c(n)\in \mathcal{A}_{c}$ denote the $n$-th secondary symbol.
In the $n$-th symbol period, the received signal at the C-Rx is given by
\footnote{Synchronization between the primary and secondary signals is critical to our system. Currently, how to achieve perfect synchronization is still an open problem. Alternatively, the envelop detector offers one way for synchronization~\cite{bharadia2015backfi}. Also, the investigations on the impact of 
asynchronous operations for tags are studied in~\cite{alevizos2023batteryless}.
For better performance of synchronization, we believe it deserves another work in the future.}
\begin{align} \label{eq-received-signal-C-Rx-1}
y(n)&=\underbrace{\sqrt{p}h_{d}s(n)}_{\text{Direct link}}+\underbrace{\sqrt{p}\bm{h}_{r}^{H}
(\bm{A}_{s}+\bm{\Phi}(c(n)))
\bm{f}s(n)}_{\text{Reflecting link}}+z(n)
\nonumber
\\ &=\sqrt{p}hs(n)+\sqrt{p}\bm{h}_{r}^{H}
\bm{\Phi}(c(n))
\bm{f}s(n)+z(n).
\end{align}
where $z(n)\sim \mathcal{CN}(0,\sigma^{2})$ is the complex Gaussian noise with zero mean
and variance $\sigma^{2}$. Note that the RIS involves both structural mode reflection and antenna mode reflection~\cite{griffin2009complete,kimionis2014increased}. In~\eqref{eq-received-signal-C-Rx-1}, $\bm{A}_{s}\in \mathbb{C}^{K \times K }$ denotes the structural mode reflection of RIS, which is related to the intrinsic physical property of conducting objects like materials and layout; $\bm{\Phi}(c(n))\in \mathbb{C}^{K \times K }$ represents the antenna mode reflection, which is a function of $c(n)$ and can be adjusted by changing the load impedance. Since the structural mode reflection is load-independent and would not be changed with the
configuration of phase shifts, we incorporate its effect into the direct link channel as
\begin{align} \label{eq-direct-link}
    h\triangleq h_{d}+\bm{h}_{r}^{H}
\bm{A}_{s}
\bm{f}.
\end{align}
The impact of the structural mode reflection on the system performance will be clearly quantified in Sec. \ref{sec;structural}. 

Additionally, our proposed design can be readily extended to the scenario where the symbol duration of the secondary signal is larger than that of the primary signal. For simplicity, we assume the primary and secondary signals have the same symbol duration in this paper.

\subsection{Conventional Modulation Scheme} \label{sec:con}
For conventional modulation schemes, such as OOK~\cite{hua2021novel} and BPSK~\cite{zhang2021reconfigurable}, the mapping rule between the secondary signal $c(n)$ and the phase-shift matrix $\bm{\Phi}(c(n))\in\mathbb{C}^{K\times K}$ is modeled as
\begin{equation}\label{eq:con}
\bm{\Phi}(c(n))= c(n)\bm{\Phi}_{\mathrm{con}},\forall c(n) \in\mathcal{A}_{c},
\end{equation}
where $\bm{\Phi}_{\mathrm{con}}\in\mathbb{C}^{K\times K}$ can be regarded as the diagonal beamforming matrix of the secondary signal $c(n)$.

Plugging \eqref{eq:con} into \eqref{eq-received-signal-C-Rx-1},
the C-Rx applies the maximum-likelihood (ML) detector to jointly recover the primary and secondary signals, which search the optimal candidate from the constellation sets with the minimum squared error (MSE), given by
\begin{align}\label{eq-ML-Detector-sc}
&\left\{\hat{s}(n),\hat{c}(n)\right\}\!=\!\underset{s(n)\in\mathcal{A}_{s},c(n)\in\mathcal{A}_{c}}{\arg \min} \nonumber \\
&  \quad \quad \quad \quad \left\| y(n)\!-\!\sqrt{p}hs(n)\!-\!\sqrt{p}\bm{h}_{r}^{H}\bm{\Phi}_{\mathrm{con}}\bm{f}s(n)c(n)\right\|^{2},
\end{align}
where $\hat{s}(n)$ and $\hat{c}(n)$ are the estimators of $s(n)$ and $c(n)$, respectively. 

From \eqref{eq-ML-Detector-sc}, $\bm{h}_{r}^{H}\bm{\Phi}_{\mathrm{con}}\bm{f}$ can be viewed as the equivalent reflecting link. It can be seen that when the reflecting link enhanced by RIS is much stronger than the direct link or the direct link is blocked by obstacles, the impact of the direct link $h$ can be ignored. In this case, the C-Rx only receives the multiplication of primary and secondary
signals, i.e., $y(n)\!=\! \sqrt{p}\bm{h}_{r}^{H}\bm{\Phi}_{\mathrm{con}}\bm{f}s(n)c(n)\!+\!z(n)$. Obviously, the C-Rx fails to detect signals due to the ambiguity
problem in this case, which is defined as

\textbf{Definition 1 (Ambiguity Problem):} \emph
{In the case of $h=0$, there exist two sets of estimators $\{\hat{s}_{1}(n),\hat{c}_{1}(n)\}$ and $\{\hat{s}_{2}(n),\hat{c}_{2}(n)\}$ that both minimize the MSE of \eqref{eq-ML-Detector-sc} when $\hat{s}_{1}(n)\hat{c}_{1}(n)$ $=\hat{s}_{2}(n)\hat{c}_{2}(n)$.}

Taking $\mathcal{A}_{s}=\{1,-1\}$ and $\mathcal{A}_{c}=\{1,-1\}$ as an example, both $\{\hat{s}_{1}(n)=1,
\hat{c}_{1}(n)=1\}$ and $\{ \hat{s}_{2}(n)=-1, \hat{c}_{2}(n)=-1\}$ are the optimal estimators to \eqref{eq-ML-Detector-sc} when $h=0$, which leads to the ambiguity for joint decoding. In practice,
when the direct link is blocked by obstacles or the reflecting link $\bm{h}_{r}^{H}\bm{\Phi}_{\mathrm{con}}\bm{f}$ is relatively much larger than the direct link, the joint detection will suffer from severe performance loss due to the ambiguity problem.
Thus, the conventional modulation schemes are not optimal for RIS-assisted SR, which motivates our work.

\section{Proposed Modulation Scheme and Its Corresponding Receiver Design} \label{sec: proposed-modulation-scheme}
In this section, to address the ambiguity problem mentioned above, we first propose a novel modulation scheme for RIS-assisted SR and then propose a novel approach for receiver design to recover both the primary and secondary signals.
\subsection{Proposed Modulation Scheme}
As mentioned above, there exists an ambiguity problem when $h=0$. Under such a scenario, if the RIS could allocate a portion of the reflection energy to purely assist the primary transmission in addition to delivering its own information, the BER performance of the primary transmission can be improved. This in turn improves the BER performance of the secondary transmission thanks to the coupling effect between them~\cite{zhang2022mutualistic}.

Accordingly, we propose a novel modulation scheme that maps the secondary signal $c(n)$ into the RIS phase-shift matrix $\bm{\Phi}(c(n))$ in the following manner
\begin{equation} \label{eq: proposed}
\bm{\Phi}(c(n))= \bm{\Phi}_{1}+c(n)\bm{\Phi}_{2}, \forall c(n) \in\mathcal{A}_{c}.
\end{equation}

In \eqref{eq: proposed},  $\bm{\Phi}_{1}=\mathrm{diag}(\phi_{1,1},\cdots,\phi_{1,K}) \in \mathbb{C}^{K\times K}$ represents the symbol-invariant component, which is used to assist the primary transmission; and $c(n)\bm{\Phi}_{2}=c(n)\mathrm{diag}(\phi_{2,1},\cdots,\phi_{2,K})\in\mathbb{C}^{K\times K}$ is the symbol-varying component, which is used for the information delivery of the secondary transmission. Here, $\bm{\Phi}_{2}$ is the passive beamforming matrix of signal $c(n)$.  Both $\bm{\Phi}_{1}$ and $\bm{\Phi}_{2}$ need to be designed, which will be detailed in Sec. V. For the special case of $\bm{\Phi}_{2}=\bm{0}_{K\times K}$, the RIS purely assists the primary transmission without its own information delivery. When $\bm{\Phi}_{1}=\bm{0}_{K\times K}$, the proposed modulation scheme degrades to its conventional counterpart given by \eqref{eq:con}.

Due to the modulus constraints of the RIS phase shifts, the proposed modulation scheme should satisfy the following constraints,
\begin{equation} \label{eq: modulus-constraint}
|\left[\bm{\Phi}_{1}+c(n)\bm{\Phi}_{2}\right]_{k,k}|\leq1, \forall c(n) \in\mathcal{A}_{c}, k=1,\cdots,K.
\end{equation}

By adopting the proposed modulation scheme,
the received signal at the C-Rx is rewritten as
\begin{align}
y(n)\!=\!\sqrt{p}(h\!+\!\bm{h}_{r}^{H}\bm{\Phi}_{1}\bm{f})s(n)\!+\!\sqrt{p}\bm{h}_{r}^{H}\bm{\Phi}_{2}\bm{f} s(n)c(n)\!+\!z(n). \nonumber
\end{align}
We can see that when $h$ is absent, the reflecting link via the symbol-invariant component $
\bm{\Phi}_{1}$ serves as a virtual direct link, which addresses the ambiguity problem given in \textbf{Definition 1}.

Note that FSK modulation schemes in backscatter communications can decode primary and secondary signals by switching the frequency of the primary signal between two frequencies~\cite{kimionis2013bistatic,fasarakis2015coherent,wang2017fm,vougioukas2018coherent}. Different from them, our proposed modulation scheme can provide an alternative way to decode primary and secondary signals without switching the frequency, and at the same time allowing the primary transmission to reap the multi-path gain from the secondary transmission.
\subsection{Receiver Design}
It can be observed that the primary and secondary signals are coupled together, which makes it quite challenging to decode them.
In this paper, instead of jointly decoding $s(n)$ and $c(n)$ directly as \cite{zhang2022mutualistic} or one after another as \cite{long2019symbiotic}, we propose a novel approach for receiver design through the use of a composite signal, which allows us to quantify both the overall and individual performance of the primary and secondary signals, given by 
\begin{align} \label{eq-x}
x(n)=\frac{1}{g}\left(h+\bm{h}_{r}^{H}\bm{\Phi}_{1}\bm{f}+\bm{h}_{r}^{H}\bm{\Phi}_{2}\bm{f}c(n)\right)s(n).
\end{align}
where $g$ is defined as the equivalent reflecting link when RIS is designed to align it, given by
\begin{align} \label{eq:reflecting-link}
    g=\sum_{k=1}^{K}|f_{k}||h_{r,k}|.
\end{align}

The composite signal $x(n)$ consists of both the primary and secondary signals, using which the received signal at the C-Rx can be rewritten as
\begin{align} \label{eq: received-composite}
    y(n)=\sqrt{p}g x(n)+z(n).
\end{align}

Combining \eqref{eq-x} and \eqref{eq: received-composite}, we can decode $s(n)$ and $c(n)$ as follows. First, we will decode the intermediate composite signal $x(n)$ from $y(n)$ using the ML detector. Then, we use the bit mapping rule to map the decoded signal $\hat{x}(n)$ into $\left\{\hat{s}(n),\hat{c}(n)\right\}$.
The above two-step procedure is summarized as follows.
\begin{align} 
&\text{Step 1}: \hat{x}(n)=\arg \min_{x(n)\in\mathcal{A}_{x}} \left\| y(n)-\sqrt{p}gx(n)\right\|^{2},\label{eq-ML-Detector-x}\\
&\text{Step 2}:\hat{x}(n)\xrightarrow{\text{bit mapping}}\left\{\hat{s}(n),\hat{c}(n)\right\},\label{eq-x-sc}
\end{align}
where $\mathcal{A}_{x}$ denotes the constellation set of $x(n)$, which can be derived using \eqref{eq-x}. 

This two-step detection provides us with useful insights regarding the phase shifts optimization of $\bm{\Phi}_{1}$ and $\bm{\Phi}_{2}$.
Based on \eqref{eq-ML-Detector-x} and \eqref{eq-x-sc}, it is observed the BER performance of $s(n)$ and $c(n)$ are related to that of $x(n)$. This means if we want to guarantee a good BER performance of $s(n)$ and $c(n)$, we first need to make sure that the signal $x(n)$ is decoded accurately. Besides, when $x(n)$ is decoded, a good bit mapping rule from $x(n)$ to $\{s(n),c(n)\}$ will reduce the number of erroneous bits for $s(n)$ and $c(n)$. Note that the design of $\bm{\Phi}_{1}$ and $\bm{\Phi}_{2}$ affect both the performance of the two steps. Therefore, they need to be carefully designed to reduce the BERs of both the primary and secondary signals. 
\begin{figure*} [t] 
	\centering  
	\includegraphics[width=5in]{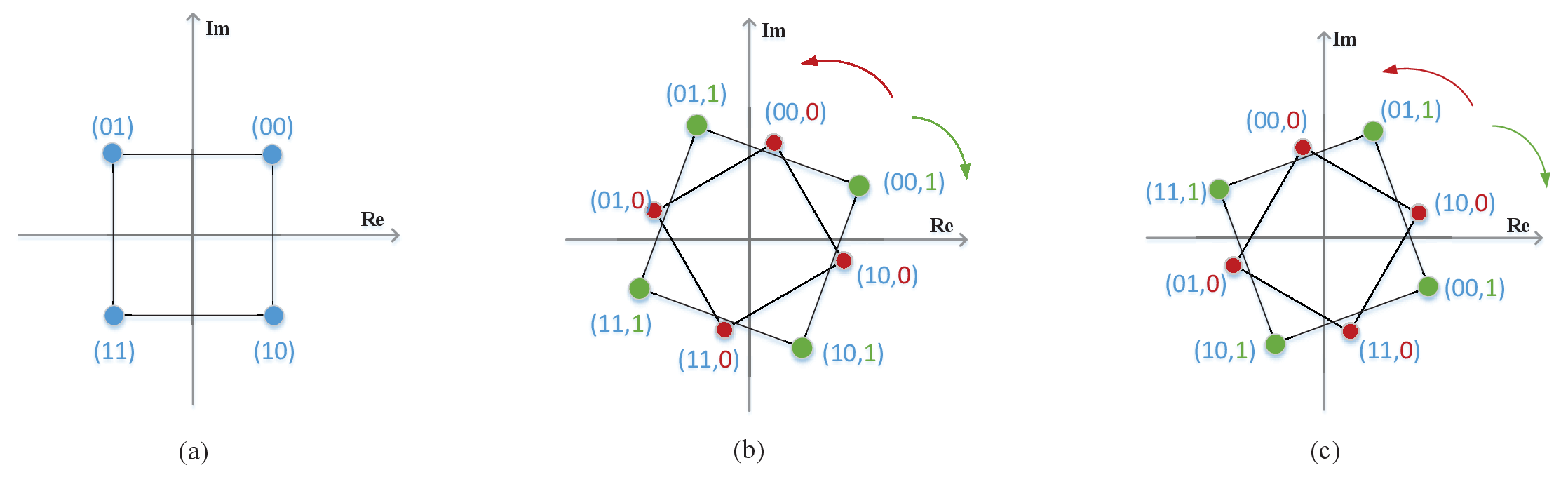}  
	\caption{
	(a) bit-mapping rule of $s$; (b) bit-mapping rule I of $x$; (c) bit-mapping rule II of $x$.} \label{fig:rotation}
\end{figure*}
\section{Design Criteria and Problem Formulation}\label{sec-modulation-design}
In this section, we introduce our design methodology for optimizing the symbol-invariant and symbol-varying components, $\bm{\Phi}_{1}$ and $\bm{\Phi}_{2}$. First, we will present our design criteria, based on which we formulate the problem to optimize $\bm{\Phi}_{1}$ and $\bm{\Phi}_{2}$.
To show our methodology more clearly, we use  $\mathcal{A}_{s} = \frac{1}{\sqrt{2}}\{1+\jmath, -1+\jmath, -1-\jmath,1-\jmath\}$, $\mathcal{A}_{c} = \{1,-1\}$ as an example.
Note that our proposed design methodology can also be extended to other higher-order modulation constellations.
For notation convenience, we normalize the transmit power $p=1$ and drop the time notation $n$ from now on.
\subsection{Design Criteria}
According to~\cite{proakis2001digital}, the minimum Euclidean distance and the Hamming distance between the adjacent constellation points affect the detection of signal $x$ and the bit mapping from $x$ to $\{s,c\}$, which corresponds to the performance of the two steps in the proposed detector. 
\subsubsection{The Minimum Euclidean Distance}
Since a larger minimum Euclidean distance leads to a better symbol error rate (SER) performance of $x$, our objective is to maximize the minimum Euclidean distance of the constellation $x$, given by
\begin{equation}
D_{\min}\triangleq\min\limits_{m \neq l}\left\{D_{m,l},m,l=1,\cdots,|\mathcal{A}_{x}|\right\}, 
\end{equation}
where $D_{m,l}$ is defined as $D_{m,l}=|x_{m}-x_{l}|$, $\forall x_{m}, x_{l}\in\mathcal{A}_{x}$. 

\subsubsection{The Hamming Distance Between the Adjacent Constellations}
This factor is very important when the BER performance is considered.
Without loss of generality, we assume the bit-mapping rule of $s$ and $c$ follow the idea of classical Gray coding, i.e., the constellation sets $\mathcal{A}_{s}$ and $\mathcal{A}_{c}$ are mapped into the bit sequences $\left\{00, 01, 11, 10\right\}$ and $\left\{1,0\right\}$, respectively. Fig. \ref{fig:rotation} (a) illustrates an example of the bit-mapping rule of $s$. Note that the composite signal $x$ has $3$ bits whose first two bits denote the primary signal $s$ and the last bit denotes the secondary signal $c$.
Given the bit-mapping rule of $s$ and $c$,
the bit-mapping rule of $x$ is affected by $\bm{\Phi}_{1}$ and $\bm{\Phi}_{2}$.
Specifically, from \eqref{eq-x}, $x$ can be viewed as the rotated versions of $s$, where the phase rotation is determined by $\angle\left(\frac{1}{g}\left(h+\bm{h}_{r}^{H}\bm{\Phi}_{1}\bm{f}+\bm{h}_{r}^{H}\bm{\Phi}_{2}\bm{f}\right)\right)$ and $\angle\left(\frac{1}{g}\left(h+\bm{h}_{r}^{H}\bm{\Phi}_{1}\bm{f}-\bm{h}_{r}^{H}\bm{\Phi}_{2}\bm{f}\right)\right)$, for $c=1$ and $c=-1$.

In summary, there are two bit-mapping rules of $x$ shown in Fig. \ref{fig:rotation} (b) and Fig. \ref{fig:rotation} (c), under the conditions $\{-\frac{\pi}{4}\leq\angle\left(\frac{1}{g}\left(h+\bm{h}_{r}^{H}\bm{\Phi}_{1}\bm{f}+\bm{h}_{r}^{H}\bm{\Phi}_{2}\bm{f}\right)\right) \leq 0, 0\leq \angle\left(\frac{1}{g}\left(h+\bm{h}_{r}^{H}\bm{\Phi}_{1}\bm{f}-\bm{h}_{r}^{H}\bm{\Phi}_{2}\bm{f}\right)\right)\leq \frac{\pi}{4}\}$ 
and
$\{-\frac{\pi}{2}\leq\angle\left(\frac{1}{g}\left(h+\bm{h}_{r}^{H}\bm{\Phi}_{1}\bm{f}+\bm{h}_{r}^{H}\bm{\Phi}_{2}\bm{f}\right)\right) \leq -\frac{\pi}{4}, \frac{\pi}{4}\leq \angle\left(\frac{1}{g}\left(h+\bm{h}_{r}^{H}\bm{\Phi}_{1}\bm{f}-\bm{h}_{r}^{H}\bm{\Phi}_{2}\bm{f}\right)\right)\leq \frac{\pi}{2}\}$, respectively. The green dots denote $\frac{1}{g} (h+\bm{h}_{r}^{H}\bm{\Phi}_{1}\bm{f}+\bm{h}_{r}^{H}\bm{\Phi}_{2}\bm{f})s$ when $c$ takes $1$, while the red dots denote $ 
\frac{1}{g} (h+\bm{h}_{r}^{H}\bm{\Phi}_{1}\bm{f}-\bm{h}_{r}^{H}\bm{\Phi}_{2}\bm{f})s$ when $c$ takes $-1$. 
Accordingly, we summarize the following properties of these two rules.

\emph{\textbf{Properties of the bit-mapping rule I in Fig. \ref{fig:rotation} (b):} }
\begin{itemize}
	\item \emph{Two adjacent 3-bit symbols in the same quadrant differ by only 1 bit.}
	\item \emph{Two adjacent 3-bit symbols in different quadrants differ by 2 bits.}
\end{itemize}

\emph{\textbf{Properties of the bit-mapping rule II in Fig. \ref{fig:rotation} (c):}}
\begin{itemize}
	\item \emph{Two adjacent 3-bit symbols in the same quadrant differ by 3 bits.}
	\item \emph{Two adjacent 3-bit symbols in different quadrants differ by 2 bits.}
\end{itemize}

Thus, we conclude that the bit-mapping rule I of Fig  \ref{fig:rotation} (b) is more appealing since the average Hamming distance between the adjacent points of Fig. \ref{fig:rotation} (b) is less than that of Fig. \ref{fig:rotation} (c).

To satisfy the bit-mapping rule I, when designing $\bm{\Phi}_{1}$ and $\bm{\Phi}_{2}$, we have the following constraints:
\begin{align}
&-\frac{\pi}{4}\leq\angle\left(\frac{1}{g}\left(h+\bm{h}_{r}^{H}\bm{\Phi}_{1}\bm{f}+\bm{h}_{r}^{H}\bm{\Phi}_{2}\bm{f}\right)\right) \leq 0, \label{eq: angle-constraint1}
\\
&\quad \ 0\leq \angle\left(\frac{1}{g}\left(h+\bm{h}_{r}^{H}\bm{\Phi}_{1}\bm{f}-\bm{h}_{r}^{H}\bm{\Phi}_{2}\bm{f}\right)\right) \leq \frac{\pi}{4}.\label{eq: angle-constraint2}
\end{align}
\subsection{Problem Formulation}
By jointly considering the criteria of the Euclidean distance and the Hamming distance, our objective is to maximize the minimum Euclidean distance of the composite signal, under the phase rotation constraints \eqref{eq: angle-constraint1} and \eqref{eq: angle-constraint2}, and the modulus constraints of RIS phase shifts, which can be formulated as the following problem
\begin{subequations}
	\begin{align}
	\underline{
 \textbf{\text{P1:}}} \quad 
	&\max \limits_{\bm{\Phi}_{1},\bm{\Phi}_{2}}\ D_{\min}\triangleq\min\limits_{m \neq l}\left\{D_{m,l},m,l=1,\cdots,|\mathcal{A}_{x}|\right\}\\
	&\ \mbox{s.t.} \ \: 
|\left[\bm{\Phi}_{1}+\bm{\Phi}_{2}\right]_{k,k}|\leq1, \forall k =1,\cdots,K, \label{eq:modulus-constraint-1} \\
	& \quad \quad |\left[\bm{\Phi}_{1}-\bm{\Phi}_{2}\right]_{k,k}|\leq1, \forall k =1,\cdots, K, \label{eq:modulus-constraint-2}
 \\ & \quad \quad
 \eqref{eq: angle-constraint1}, \eqref{eq: angle-constraint2}. \nonumber
\end{align}
\end{subequations}
where \eqref{eq:modulus-constraint-1} and \eqref{eq:modulus-constraint-2} are the modulus constraints from \eqref{eq: modulus-constraint} under different values of $c$.
One can observe that it is challenging to solve \textbf{P1} since the phase rotation constraints \eqref{eq: angle-constraint1} and \eqref{eq: angle-constraint2}) are non-convex, and the objective function $D_{\min}$ is not jointly concave with respect to $\bm{\Phi}_{1}$ and $\bm{\Phi}_{2}$.
Generally speaking, there is no standard optimization method to directly obtain the optimal solution\footnote{Here, we assume perfect instantaneous CSI is available to facilitate our design and analysis. The channel estimation approaches can be found in~\cite{swindlehurst2022channel}. Besides, considering the CSI estimation errors in practice, we will examine its impact on the BER performance in Sec. \ref{sec;structural}. Moreover, optimization based on instantaneous CSI requires more overhead, and therefore statistical CSI-based beamforming or blind beamforming is worth future investigation.}.
\section{Optimization Algorithm for \textbf{P1}} \label{sec: geometrical-analysis}
In this section, we aim to solve \textbf{P1}. First, we will make some transformations to \textbf{P1} to simplify its phase-shift matrix design. Then, the geometrical analysis method is employed to derive the closed-form solution.
\subsection{Problem Transformation}
First, we analyze the problem \textbf{P1} by exploiting the structure of its objective function, which will help us to simplify the design of the phase-shift matrix, shown in the following Proposition.
\begin{prop} \label{PropWeightedChannel}
	The phase-shift matrix of RIS, $\bm{\Phi}(c)=\bm{\Phi}_{1}+c\bm{\Phi}_{2}$, can be equivalently rewritten as the following form
 \begin{align} \label{eq: equivalent-form}
     \bm{\Phi}(c)=\bm{\Phi}_{1}+c\bm{\Phi}_{2}=\bm{\Phi}(\alpha+\beta c),
 \end{align}
  where $\bm{\Phi}_{1}=\alpha \bm{\Phi}$ and $\bm{\Phi}_{2}=\beta \bm{\Phi}$; $\alpha\in\mathbb{R}$ and $\beta\in \mathbb{C}$ represent the weighted parameters; $\bm{\Phi}$ denotes the common phase-shift matrix shared by $\bm{\Phi}_{1}$ and $\bm{\Phi}_{2}$, given by $\bm{\Phi}=e^{\jmath \angle (h)}\mathrm{diag}[e^{- \jmath \angle (f_{1}h_{r,1}^{H})},\cdots,e^{- \jmath \angle (f_{K}h_{r,K}^{H})}]$, and $\angle(\cdot)$ denotes the phase of the argument.
\end{prop}
\begin{IEEEproof}
	Please refer to Appendix \ref{Appendix-prop1}.
\end{IEEEproof}

Proposition \ref{PropWeightedChannel} shows the following insights. With the equivalent form shown in \eqref{eq: equivalent-form}, the phase-shift matrix optimization boils down to optimizing the two weighted parameters, i.e., $\alpha$ and $\beta$. Obviously, the common phase-shift matrix $\bm{\Phi}$ is designed to align the reflecting link (PTx--RIS--C-Rx link) such that the signal strength of the reflecting link can be maximized, and the composite signal becomes $x=(\frac{h}{g}+\alpha+\beta c)s$ in this case.
Next, we only need to adjust the parameters $\alpha$ and $\beta$ to control the values of $\bm{\Phi}_{1}$ and $\bm{\Phi}_{2}$.


With the above transformations, the problem \textbf{P1} can be recast as
\begin{subequations}
	\begin{align}
	\underline{\textbf{P1}^{'}\textbf{:}} \quad 
	&\max \limits_{\alpha,\beta}\ D_{\min}\triangleq\min\limits_{m \neq l}\left\{D_{m,l},m,l=1,\cdots,|\mathcal{A}_{x}|\right\}\\
	&\ \mbox{s.t.} \ \: \mathrm{C1}: -\frac{\pi}{4}\leq\angle\left(\frac{h}{g}+\alpha+\beta\right) \leq 0, 
 \\
	& \quad \quad \mathrm{C2}:  0\leq \angle\left(\frac{h}{g}+\alpha-\beta\right) \leq \frac{\pi}{4}, \\
	& \quad \quad\mathrm{C3}: |\alpha+\beta|\leq1, \\
	& \quad \quad\mathrm{C4}: |\alpha-\beta|\leq1.
	\end{align}
\end{subequations}


Next, we will derive the optimal solution of $\alpha^{\star}$ and $\beta^{\star}$. To do so, we first optimize $\beta$ under a given $\widetilde{\alpha}$ and derive the closed-form expression of $\beta^*(\widetilde{\alpha})$ and the minimum Euclidean distance $D_{\min}^*(\widetilde{\alpha})$. After that, we further optimize $\alpha$ by maximizing $D_{\min}^*(\widetilde{\alpha})$, and then obtain the optimal solution $(\alpha^{\star}, \beta^{\star})$. For ease of analysis, we let $\angle (h/g)=0$. For the case of $\angle(h/g)\neq0$, we can directly multiply the optimal $\alpha^{\star}$ and $\beta^{\star}$ by $e^{\jmath\angle(h/g)}$. 

\subsection{Optimization of $\beta$ under a given $\widetilde{\alpha}$}
For convenience, \textbf{P1}$^{'}$ under a given $\widetilde{\alpha}$ is denoted as \textbf{P2}. Although \textbf{P2} is a single variable problem with respect to $\beta$, it is still difficult to obtain the optimal solution due to the following reasons. First, the variable $\beta$ is constrained by the constraints $\mathrm{C1}$-$\mathrm{C4}$, where the feasible region is not explicitly characterized. Second, the objective function is the minimum one among all the Euclidean distances, which is determined by $\widetilde{\alpha}$ and $\beta$. However, it is not clear which pair of the Euclidean distance set is the minimum. 

To solve \textbf{P2}, we first need to rewrite the constraints $\mathrm{C1}$-$\mathrm{C4}$ to explicitly characterize the feasible region of $\beta$. Let $\beta=|\beta|e^{\jmath\theta}$, where $\theta$ denotes the phase of $\beta$. Specifically, we can recast the constraints $\mathrm{C1}$ and $\mathrm{C2}$ as the following constraint $\mathrm{\overline{C2}}$ based on Appendix \ref{Appdendix: rotate}.
\begin{align} \label{eq:phase-beta}
&\mathrm{\overline{C2}}:~\beta \in \left\{\theta, |\beta| ~ \Big |-\pi\leq\theta\leq0, 0\leq|\beta|\leq  \right.
\nonumber
\\& \quad \quad \quad \quad \quad  \phantom{=\;\;}
	\left.
\frac{1}{\cos(\theta)(-1)^{\mathbb{I}_{1}(\theta)}-\sin(\theta)}\left|\frac{h}{g}+\widetilde{\alpha}\right| \right\},
\end{align}
where $\mathbb{I}_{1}(\theta)$ is an indicator function (i.e., $\mathbb{I}_{1}(\theta)=1$, if $-\pi\leq\theta\leq-\frac{1}{2}\pi$; $\mathbb{I}_{1}(\theta)=0$, if $-\frac{1}{2}\pi<\theta\leq0$ ).

In addition, the constraints $\mathrm{C3}$ and $\mathrm{C4}$ can be transformed into the following constraint $\mathrm{\overline{C4}}$ according to Appendix \ref{Appendex_modulous-one}.
\begin{align} \label{eq: c4-ba}
&\mathrm{\overline{C4}}:
\beta \in \left\{
\theta, |\beta|~ \Big | -\pi \leq \theta\leq \pi, 0\leq|\beta|\leq \right.
\nonumber
\\
& \quad \quad \quad   \phantom{=\;\;}
	\left.
\sqrt{1-|\widetilde{\alpha}|^2\sin^2(\theta)}-|\widetilde{\alpha}|\cos(\theta)(-1)^{\mathbb{I}_{2}(\theta)}
\right\},
\end{align}
where $\mathbb{I}_{2}(\theta)$ is an indicator function ( $\mathbb{I}_{2}(\theta)=1$, if $\frac{1}{2}\pi<\theta\leq\frac{3}{2}\pi$; $\mathbb{I}_{2}(\theta)=0$, if $-\frac{1}{2}\pi<\theta\leq\frac{1}{2}\pi$).

Then, with the transformed $\mathrm{\overline{C2}}$, $\mathrm{\overline{C4}}$, \textbf{P2} can be recast as \textbf{P3}
\begin{subequations}
	\begin{align}
	\underline{\textbf{\text{P3:}}} \quad 
	&\max \limits_{\beta} \ D_{\min}\triangleq\min\limits_{m \neq l}\left\{D_{m,l},m,l=1,\cdots,|\mathcal{A}_{x}|\right\} \\
	&\ \mbox{s.t.} \ \mathrm{\overline{C2},\overline{C4}}.
	\end{align}
\end{subequations}

Although simplified, it is still difficult to solve \textbf{P3} in its current form. To solve it, we present an useful Lemma to determine the minimum Euclidean distance for a given $\widetilde{\alpha}$.
\begin{lem} \label{lem-D-min-general}
	Let $\beta=|\beta|e^{\jmath\theta}$, where $\theta$ denotes the phase of $\beta$.
	The minimum Euclidean distance of the constellations $x$ under a given $\widetilde{\alpha}$, i.e., $D_{\min}(\beta | \widetilde{\alpha})$, is expressed as
	\begin{align} \nonumber
	   D_{\min}(\beta | \widetilde{\alpha})\!=\!
	\begin{cases}
	D_{1}\!=\!2|\beta|,& \beta\!\in\!\mathbb{D}_{1},\\
	D_{2}\!=\!\sqrt{2}|\frac{h}{g}+\widetilde{\alpha}-\beta|,&\beta\!\in\!\mathbb{D}_{2},  
 \\
	D_{3}\!=\!\sqrt{2}|\frac{h}{g}+\widetilde{\alpha}-\jmath\beta|, & \beta\!\in\!\mathbb{D}_{3}, 
 \\
	D_{4}\!=\!\sqrt{2}|\frac{h}{g}+\widetilde{\alpha}+\beta|,& \beta\!\in\!\mathbb{D}_{4}, \\
	D_{5}\!=\!2|\frac{h}{g}+\widetilde{\alpha}| ,&\beta\!\in\!\mathbb{D}_{5},
	\end{cases} 
	\end{align}
	where
 \begin{align}
     &\mathbb{D}_{1}\!\triangleq\!\{\theta,|\beta| \big| -\pi\leq\theta\leq0,0\leq|\beta|\leq|\beta_{5}(\theta)|\},
     \\
     &\mathbb{D}_{2}\!\triangleq\!\{\theta,|\beta| \big |-\frac{\pi}{4}\leq\theta\leq0,|\beta_{1}(\theta)|\leq|\beta|\leq |\beta_{2}(\theta)|\},
     \\
     & \mathbb{D}_{3}\!\triangleq\! \{\theta,|\beta| \big |-\frac{3\pi}{4}\!\leq\!\theta\!\leq\!-\frac{\pi}{4},|\beta_{3}(\theta)|\!\leq\!|\beta|\!\leq \!|\beta_{4}(\theta)|\},
     \\
     & \mathbb{D}_{4}\!\triangleq\! \{\theta,|\beta| \big |-\pi\!\leq\!\theta\!\leq\!-\frac{3}{4}\pi,|\beta_{2}(\theta)|\!\leq\!|\beta|\!\leq\! |\beta_{1}(\theta)|\},
     \\
     & \mathbb{D}_{5}\!\triangleq\!\{\theta,|\beta| \big| -\pi\leq\theta\leq0,|\beta|\geq|\beta_{6}(\theta)|\},
 \end{align}
and  $|\beta_{1}(\theta)|=|\frac{h}{g}+\widetilde{\alpha}|\left(-\cos(\theta)+\sqrt{1+\cos^{2}(\theta)}\right)$,
	$|\beta_{2}(\theta)| =|\frac{h}{g}+\widetilde{\alpha}|\left(\cos(\theta)+\sqrt{1+\cos^{2}(\theta)}\right)$, $|\beta_{3}(\theta)|=|\frac{h}{g}+\widetilde{\alpha}|\left(\sin(\theta)+\sqrt{1+\sin^{2}(\theta)}\right)$,
	$|\beta_{4}(\theta)|=|\frac{h}{g}+\widetilde{\alpha}|\left(-\sin(\theta)+\sqrt{1+\sin^{2}(\theta)}\right)$, 
	$|\beta_{5}(\theta)|=\mathbb{I}_{3}(\theta)|\beta_{1}(\theta)|+\mathbb{I}_{4}(\theta)|\beta_{3}(\theta)|+\mathbb{I}_{5}(\theta)|\beta_{2}(\theta)|$,
	$|\beta_{6}(\theta)|=\mathbb{I}_{3}(\theta)|\beta_{2}(\theta)|+\mathbb{I}_{4}(\theta)|\beta_{4}(\theta)|+\mathbb{I}_{5}(\theta)|\beta_{1}(\theta)|$,  $\mathbb{I}_{3}(\theta)$, $\mathbb{I}_{4}(\theta)$, and $\mathbb{I}_{5}(\theta)$ are indicator functions ( $\mathbb{I}_{3}(\theta)=1$, if $-\frac{1}{4}\pi<\theta\leq0$; $\mathbb{I}_{3}(\theta)=0$, otherwise),  ($\mathbb{I}_{4}(\theta)=1$, if $-\frac{3}{4}\pi<\theta\leq-\frac{1}{4}\pi$; $\mathbb{I}_{4}(\theta)=0$, otherwise), ($\mathbb{I}_{5}(\theta)=1$, if $-\pi<\theta\leq-\frac{3}{4}\pi$; $\mathbb{I}_{5}(\theta)=0$, otherwise).
\end{lem}
\begin{IEEEproof}
	Please refer to Appendix \ref{Appendix-D-min-general} for details.
\end{IEEEproof}

Lemma \ref{lem-D-min-general} divides the region of $\beta$ into five distinct regions, in which the minimum Euclidean distance is determined. 
With Lemma \ref{lem-D-min-general}, we can obtain five solutions of $\beta$ to \textbf{P3} under the five regions, respectively. Then, we choose the one that maximizes the minimum Euclidean distance, i.e., $D_{\min}(\beta|\widetilde{\alpha})$. Specifically, we can recast the \textbf{P3} as five subproblems, denoted by \textbf{P3-k}, where $k = 1,\cdots,5$.
\begin{subequations}
	\begin{align}
	\underline{\textbf{\text{P3-k:}}} \quad 
	&\max \limits_{\beta} \ D_{\min}\triangleq D_{k} \\
	&\ \mbox{s.t.} \ \mathrm{\overline{C2},\overline{C4}}, \mathrm{C5}: \beta\in\mathbb{D}_{k}.
	\end{align}
\end{subequations}

For \textbf{P3-k}, the constraints $\mathrm{\overline{C2},\overline{C4},C5}$ explicitly show the feasible region of $\beta$. 
We can sketch these regions and take their intersection as the feasible region of $\beta$. Then the optimal $\beta_{k}^{*}$ that maximizes the objective function $D_{k}$ can be obtained through geometrical analysis (Please refer to Appendix \ref{AppendexD}).
Finally, by comparing the five solutions of $\beta$, the optimal $\beta^{*}$ to \textbf{P3} is given by 
\begin{equation} \label{eq:chooes-beta}
\beta^{*}=\arg\max\limits_{\beta=\beta_{1}^{*},\cdots,\beta_{5}^{*}} D_{\min}(\beta\big|\widetilde{\alpha}).
\end{equation}

Next, by considering different values of $\widetilde{\alpha}$, we summarize the optimal solutions $\beta^{*}$ to  \textbf{P3} in the following Theorem.
\begin{thm} \label{Theorem-withD-beta}
	Given $\widetilde{\alpha}$, the optimal $\beta^{*}$ to \textbf{P3} and the maximized minimum Euclidean distance are given by
	\begin{equation} \label{eq: beta-alpha}
	\beta^{*}(\widetilde{\alpha}) = 
	\begin{cases}
	\frac{\sqrt{6}-\sqrt{2}}{2}|\frac{h}{g}+\widetilde{\alpha}|e^{-\jmath\frac{\pi}{4}},&\ 0\leq \widetilde{\alpha} \leq \widetilde{\alpha}_{1},
	\\
	|\beta_{1}(\theta_{1})|e^{\jmath\theta_{1}},&\ \widetilde{\alpha}_{1} \leq \widetilde{\alpha} \leq 
	\widetilde{\alpha}_{2},
	\\
	\sqrt{1-\widetilde{\alpha}^2}e^{-\jmath\frac{1}{2}\pi},&\ \widetilde{\alpha}_{2} \leq \widetilde{\alpha} \leq 1
	.
	\end{cases}
	\end{equation}
\begin{small}
    \begin{align} \label{eq: distance-alpha}
	D_{\mathrm{min}}^{*}(\widetilde{\alpha})=\begin{cases}
	(\sqrt{6}-\sqrt{2})|\frac{h}{g}+\widetilde{\alpha}|, & \ 0\leq\widetilde{\alpha}\leq\widetilde{\alpha}_{1},\\
	2|\beta_{1}(\theta_{1})|,&\ \widetilde{\alpha}_{1}\leq \widetilde{\alpha} \leq 
	\widetilde{\alpha}_{2}, \\
	2\sqrt{1-\widetilde{\alpha}^2},& \ \widetilde{\alpha}_{2}\leq \widetilde{\alpha} \leq 1.
	\end{cases}
	\end{align}
\end{small}
	where 
 \begin{align}
     &\widetilde{\alpha}_{1}\!=\!0.25(\sqrt{8\!-\!0.54\!\left|h/g\right|^{2}}\!-\!1.27\!\left|h/g\right|),
     \\
     &\widetilde{\alpha}_{2}\!=\!0.43\left(-0.34\!\left|h/g\right|\!+\!\sqrt{-0.68\!\left|h/g\right|^{2}\!+\!4.7}\right),
     \\
     & \theta_{1}\!=\!\mathrm{root}\left(\sqrt{1\!-\!\widetilde{\alpha}^2\sin^2(\theta)}\!-\!\widetilde{\alpha}\cos(\theta)\!=\!|\beta_{3}(\theta)|\right),
 \end{align}
 and $|\beta_{1}(\theta)|$ is given in Lemma \ref{lem-D-min-general}, and $\mathrm{root}(\cdot)$ denotes the root of the equation.
\end{thm}
\begin{IEEEproof}
	Please refer to Appendix \ref{AppendexD} for details.
\end{IEEEproof}

Theorem \ref{Theorem-withD-beta} characterizes the expressions of optimal $\beta$ and the minimum Euclidean distance as functions of $\widetilde{\alpha}$, respectively.
In the following, we only need to analyze the monotonicity of $D_{\min}^{*}(\widetilde{\alpha})$ with respect to $\widetilde{\alpha}$ and find the optimal $\widetilde{\alpha}$ under which $D_{\min}^{*}(\widetilde{\alpha})$ is maximized.
\subsection{Optimization of $\alpha$} \label{section: optimization-alpha}
From \eqref{eq: distance-alpha}, it is observed that $D_{\mathrm{min}}^{*}(\widetilde{\alpha})$ monotonically increases over $\widetilde{\alpha}$ when $\widetilde{\alpha}\in[0,\widetilde{\alpha}_{1}]$, and monotonically decreases when $\widetilde{\alpha}\in[\widetilde{\alpha}_{2},1]$. Besides, it can be proved that $D_{\mathrm{min}}^{*}(\widetilde{\alpha})$ first decreases and then increases when $\widetilde{\alpha}\in[\widetilde{\alpha}_{1},\widetilde{\alpha}_{2}]$. In summary, we conclude that $D_{\mathrm{min}}^{*}(\widetilde{\alpha})$ attains its maximum at either $\widetilde{\alpha}_{1}$ or $\widetilde{\alpha}_{2}$. Then we choose the optimal $\alpha^{\star}=\arg \max\limits_{\widetilde{\alpha}=\widetilde{\alpha}_{1},\widetilde{\alpha}_{2}} D_{\min}^{*}(\widetilde{\alpha})$, and the optimal $\beta^{\star}$ is obtained using \eqref{eq: beta-alpha}. Note that whether $\widetilde{\alpha}_{1}$ or $\widetilde{\alpha}_{2}$ determines $D_{\min}*(\widetilde{\alpha})$ depends on the value of $|h/g|$. Thus, we give the optimal solution of $\alpha$ and $\beta$ to \textbf{P1}$^{'}$ in the following theorem under different values of $|h/g|$. 
\begin{thm} \label{thm: optimal-a-b}
	The optimal solution of $\alpha$ and $\beta$ to \textbf{P1}$^{'}$ is given by
	
	\begin{itemize}
		\item Case 1: $|h/g|=0$. The optimal $x$ forms a standard 8PSK constellation, and the optimal solution is given by
		\begin{equation} 
		\alpha^{\star}=\cos(\pi/8),\beta^{\star}=-\jmath\sin(\pi/8).
		\end{equation}
		\item Case 2: $0\!<\!|h/g|\!<\!1$. The optimal $x$ forms a scaled 8PSK scaled by $\left(1\!+\!\tan^{2}(\pi/8)\right)(|h/g|\!+\!|\alpha^{\star}|)^2$,
		where 
		\begin{equation}
		\alpha^{\star}=\widetilde{\alpha}_{2}, \beta^{\star}=\sqrt{1-|\alpha^{\star}|^2}e^{-\jmath\frac{1}{2}\pi},
		\end{equation}
		and $\widetilde{\alpha}_{2}$ is given in Theorem \ref{Theorem-withD-beta}.
		\item Case 3: $1\leq|h/g|< \frac{2}{\sqrt{6}-\sqrt{2}}$. The optimal $x$ forms a $8$QAM-like constellation, and the optimal solution is given by
		\begin{align} \small
		\alpha^{\star}\!=\!\widetilde{\alpha}_{1},  \beta^{\star}\!=\!|\beta_{1}(\theta_{1}^{\star})|e^{\jmath \theta_{1}^{\star}},
		\end{align}
		where $\widetilde{\alpha}_{1}$ is given in Theorem \ref{Theorem-withD-beta}, and $\theta_{1}^{\star}$ is given by Theorem \ref{Theorem-withD-beta} by replacing $\widetilde{\alpha}$ with $\alpha^{\star}$, and $|\beta_{1}(\theta_{1}^{\star})|$ is given in Lemma \ref{lem-D-min-general}.
		\item Case 4: $\frac{2}{\sqrt{6}-\sqrt{2}}\leq|h/g|<\sqrt{2}+1$. The optimal $\alpha^{\star}=0$, $|\beta^{\star}|=1$, $\angle \beta^{\star}\in\left(\theta_{2}^{\star},\theta_{3}^{\star}\right)$,
		 where 
		\begin{align}
		\theta_{2}^{\star} &= \mathrm{root}\left(|\beta_{1}(\theta)|=1\right), \forall\theta\in(-\frac{1}{4}\pi,0).\\
  \theta_{3}^{\star}& = \mathrm{root}\left(|\beta_{3}(\theta)|=1\right), \forall\theta\in(-\frac{1}{2}\pi,-\frac{1}{4}\pi).
		\end{align}
		\item Case 5: $ |h/g|\geq\sqrt{2}+1$. The optimal $\alpha^{\star}=0$, $|\beta^{\star}|=1$, $\angle \beta^{\star}\in(-\pi,0)$.
	\end{itemize}
\end{thm}

Theorem \ref{thm: optimal-a-b} shows that the optimal design is related to the channel strength ratio of the direct link to the reflecting link, i.e., $|h/g|$, when RIS is designed to align with them. Particularly, when the direct link is blocked or the reflecting link enhanced by RIS is much stronger than the direct link, the optimal design can address the ambiguity problem of the conventional modulation scheme since the introduction of $\alpha$ can be viewed as a virtual direct link. 
Generally speaking, the direct link is stronger than the reflecting link, apart from some special cases where the direct link is heavily blocked by multiple obstacles or the reflecting link is enhanced by increasing the number of reflecting elements of the RIS~\cite{pei2021ris,ren2022configuring}.
When the direct link is relatively weak, RIS will allocate more energy to the symbol-invariant component (i.e., $\alpha$).
As the strength ratio increases, the performance of the primary transmission is gradually dominated by the strength of the direct link. Then, RIS will reduce the energy of the symbol-invariant component and increase the energy of the symbol-varying component to enhance the performance of the secondary transmission. When the direct link is strong enough, there is no need to allocate energy to the symbol-invariant component, namely, $\alpha=0$, while the energy of the symbol-varying component is set to $|\beta|=1$. 

Similar to those works of RIS~\cite{hua2021novel}, during the channel coherence time, the PTx first estimates the channels, based on which the phase shifts (i.e., $\bm{\Phi}$, $\alpha$, and $\beta$) are optimized at the PTx. Then, the PTx sends the information of optimized parameters to the RIS controller, and then changes its phase shifts~\cite{griffin2009complete}, where the detailed circuits can be found in~\cite{kimionis2016pulse}.
Moreover, our proposed design methodology can be readily extended to other modulation constellations. In cases of high-order modulations, the geometrical methods proposed in our work are still applicable but it will be very challenging to obtain the closed-form solutions. In this case, the classical convex/non-convex optimization methods can be developed to obtain the numerical solutions.
\section{Performance Analysis} \label{sec-performance-analysis}
To evaluate the performance of the proposed modulation scheme, in this section, we analyze the theoretical BER performance. 
By using the ML detector in \eqref{eq-ML-Detector-x} and \eqref{eq-x-sc}, the BER of $x$, $s$, and $c$ are given by
\begin{small}
    \begin{equation}\label{eq:BER-x}
    P_{x}\!=\! \sum_{i=1}^{|\mathcal{A}_{x}|}\sum_{j=1,j\neq i}^{|\mathcal{A}_{x}|}P(\hat{x}\!=\!x_{j} \big| x \!=\! x_{i})P(x\! = \!x_{i})\frac{d(x_{i}\rightarrow x_{j})}{\log(|\mathcal{A}_{x}|)},
\end{equation}
\end{small}
\begin{small}
    \begin{equation} \label{eq:BER-s}
P_{s}= \sum_{i=1}^{|\mathcal{A}_{x}|}\sum_{j=1,j\neq i}^{|\mathcal{A}_{x}|}P(\hat{x}=x_{j} \big| x = x_{i})P(x = x_{i})\frac{d(s_{i}\rightarrow s_{j})}{\log(|\mathcal{A}_{s}|)},
\end{equation}
\end{small}
\begin{small}
    \begin{equation} \label{eq:BER-c}
P_{c}= \sum_{i=1}^{|\mathcal{A}_{x}|}\sum_{j=1,j\neq i}^{|\mathcal{A}_{x}|}P(\hat{x}=x_{j} \big| x = x_{i})P(x = x_{i})\frac{d(c_{i}\rightarrow c_{j})}{\log(|\mathcal{A}_{c}|)},
\end{equation}
\end{small}
where $|\mathcal{A}_{x}|$, $|\mathcal{A}_{s}|$, and $|\mathcal{A}_{c}|$ denote the cardinality of the sets $\mathcal{A}_{x}$, $\mathcal{A}_{s}$, and $\mathcal{A}_{c}$, respectively; $d(x_{i}\rightarrow x_{j})$, $d(s_{i}\rightarrow s_{j})$, and $d(c_{i}\rightarrow c_{j})$ denote the Hamming distance between $x_{i}$ and $x_{j}$, $s_{i}$ and $s_{j}$, $c_{i}$ and $c_{j}$, respectively. Note that a unique set of $\left\{s_{i},c_{i}\right\}$ can be obtained based on the known $x_{i}$ according to \eqref{eq-x}. Besides, $P(x=x_{i})$ denotes the probability that $s_{i}$ and $c_{i}$ are the transmitted symbols of the PTx and the STx, 
and we assume that $P(x=x_{i})=\frac{1}{|\mathcal{A}_{x}|}$, $\forall i =1,\cdots,|\mathcal{A}_{x}|$.
$P(\hat{x}=x_{j} \big| x = x_{i})$ denotes the probability of selecting $\hat{x}=x_{j}$ as the true point when $x=x_{i}$ is transmitted, given by
\begin{equation} \label{eq:conditional-probability}
    P(\hat{x}=x_{j} \big| x = x_{i})=P_{i,j}= \int_{y\in\mathcal{B}_{j}}p(y|x=x_{i})dy,
\end{equation}
where $\mathcal{B}_{j}$ denotes the decision region of symbol $x_{j}$, and $p(y|x=x_{i})$ is the conditional density function of $y$ given $x=x_{i}$.
As aforementioned, the composite signal, $x$, adopts the bit-mapping rule I shown in Fig. \ref{fig:rotation}, then we have the following BER relationship.
\begin{prop} \label{prop: BER relationship}
	The BER relationship among signals $x$, $s$, and $c$ is shown as $P_{x}=\frac{2}{3}P_{s}+\frac{1}{3}P_{c}$.
\end{prop}
\begin{IEEEproof}
	 Proposition \ref{prop: BER relationship} can be readily proved by expanding \eqref{eq:BER-x}, \eqref{eq:BER-s}, and \eqref{eq:BER-c}.
\end{IEEEproof}

With Proposition \ref{prop: BER relationship},
we aim to analyze the instantaneous BER by considering three special scenarios for theoretical analysis, including $|h/g|=0$, $|h/g|>>\sqrt{2}+1$, and other values of $|h/g|$.
Besides, in simulations, we will show the average performance by averaging the obtained instantaneous BER over independent channel realizations. The theoretical derivation of the average BER is left for our future work due to space limitation.
\subsection{Scenario 1: $|h/g|=0$}
In this scenario, the PTx can only communicate with the C-Rx via the reflecting link. In terms of Theorem \ref{thm: optimal-a-b}, the optimized composite signal $x$ forms a standard $8$PSK constellation with $\alpha^{\star}=\cos(\pi/8)$, $\beta^{\star}=-\jmath\sin(\pi/8)$. In this case, the decision region $\mathcal{B}_{i}$ can be described as $\mathcal{B}_{i}=\{\phi~\big |~ -\frac{\pi}{8}+\frac{\pi}{4}(i-1)<\phi\leq\frac{\pi}{8}+\frac{\pi}{4}(i-1)\}$, $i=1,\cdots,8$. By calculating the conditional probability $P(\hat{x}=x_{j} \big| x = x_{i})$, and then substituting into \eqref{eq:BER-x}, \eqref{eq:BER-s}, and \eqref{eq:BER-c}, we obtain the semi-closed form BER expressions in the following Theorem.
\begin{thm} \label{thm: BER-x}
	Let $\widetilde{\gamma_{b}}=|g|^{2}/\sigma^{2}$ denote the instantaneous SNR of the reflecting link, $P_{i,j}$ denote the probability $P(\hat{x}=x_{j} \big| x = x_{i})$. Then
	the BER of $x$, $s$, and $c$ under $|h/g|=0$ are given by
	\begin{align}
	P_{x}&= \frac{1}{3}\left(P_{12}\!+\!P_{14}\!+\!P_{15}\right)\!+\!\frac{2}{3}\left(P_{13}\!+\!P_{16}\!+\!P_{18}\right)\!+\!P_{17}, \label{eq: BERx} 
	\\
	P_{s}& = \frac{1}{2}\left(P_{12}+P_{14}+P_{16}+P_{18}\right)+P_{13}+P_{17}, \label{eq: BERs} 
	\\
	P_{c}& = P_{15}+P_{16}+P_{17}+P_{18},\label{eq: BERc} 
	\end{align}
	where
	\begin{align}
	&P_{15}=P_{18}=\frac{1}{8}e^{-\widetilde{\gamma_{b}}}+\frac{1}{2}\sqrt{\frac{\widetilde{\gamma_{b}}}{\pi}}\int_{\frac{\pi}{8}}^{\frac{3\pi}{8}}\widetilde{f}(\phi)d\phi,  \\ & P_{12}=P_{14}=\frac{1}{8}e^{-\widetilde{\gamma_{b}}}+\frac{1}{2}\sqrt{\frac{\widetilde{\gamma_{b}}}{\pi}}\int_{\frac{3\pi}{8}}^{\frac{5\pi}{8}}\widetilde{f}(\phi)d\phi, \label{eq:P1}\\
	&P_{16}=P_{17}=\frac{1}{8}e^{-\widetilde{\gamma_{b}}}+\frac{1}{2}\sqrt{\frac{\widetilde{\gamma_{b}}}{\pi}}\int_{\frac{5\pi}{8}}^{\frac{7\pi}{8}}\widetilde{f}(\phi)d\phi,\\
 &P_{13}=\frac{1}{8}e^{-\widetilde{\gamma_{b}}}+\sqrt{\frac{\widetilde{\gamma_{b}}}{\pi}}\int_{\frac{7\pi}{8}}^{\pi}\widetilde{f}(\phi)d\phi,\label{eq: P2}\\
	& \widetilde{f}(\phi)=e^{-\widetilde{\gamma_{b}}\sin^{2}(\phi)}\cos(\phi) \left(1+\mathrm{erf}(\sqrt{\widetilde{\gamma_{b}}}\cos(\phi))\right).
	\end{align}
\end{thm}
\begin{IEEEproof}
	Please refer to Appendix \ref{AppendexE} for details.
\end{IEEEproof}
\subsection{Scenario 2: $|h/g|>>\sqrt{2}+1$}
Note that $\sqrt{2}+1$ is a threshold of \emph{Case 5} in Theorem \ref{thm: optimal-a-b}. For the scenario with $|h/g|>>\sqrt{2}+1$, the direct link is relatively strong as compared to the reflecting link, and we have $\alpha^{\star}=0$, $|\beta^{\star}|=1$. The received signal can be rewritten as $y=hs+gsc+z$, where $g$ is defined in \eqref{eq:reflecting-link}. In this case, the detection of $s$ is dominated by the strong direct link, and the detection of $c$ shows a negligible impact on that of $s$. Thus, $P_{s}$ approaches  $\mathcal{Q}(|h|/\sigma)$, where $\mathcal{Q}(t)\triangleq\int_{t}^{\infty}\frac{1}{\sqrt{2\pi}}e^{-\frac{1}{2}\eta^{2}}d\eta$ denotes the complementary distribution function of the standard Gaussian.
Since the $\mathcal{Q}$-function is a monotonically decreasing function, it is observed that a larger $|h|$ will lead to a better BER performance of $P_{s}$.
Besides, when $|h/g|\rightarrow \infty$, $P_{s}$ will approach $0$. This means when the signal $s$ is decoded almost perfectly, $P_{c}$ will approach the lower bound $\mathcal{Q}(\sqrt{2}|g|/\sigma)$. By leveraging Proposition \ref{prop: BER relationship}, Theorem \ref{BER:x-s-c-scenario2} follows.
\begin{thm}\label{BER:x-s-c-scenario2}
    Let $\widetilde{\gamma_{d}}=|h|^2/\sigma^2$, $\widetilde{\gamma_{b}}=|g|^2/\sigma^2$ denote the instantaneous SNR of the direct link and the reflecting link, respectively. Then the BER of $x$, $s$, and $c$ under $|h/g|>>\sqrt{2}+1$ are given as follows.
        \begin{align}
        &P_{x}\approx \frac{2}{3}\mathcal{Q}(\sqrt{\widetilde{\gamma_{d}}})+\frac{1}{3}\mathcal{Q}(\sqrt{2\widetilde{\gamma_{b}}}), \\ 
        & P_{s}\approx \mathcal{Q}(\sqrt{\widetilde{\gamma_{d}}}), \quad
        P_{c}\approx \mathcal{Q}(\sqrt{2\widetilde{\gamma_{b}}}),
    \end{align}
\end{thm}
Theorem \ref{BER:x-s-c-scenario2} shows that when $|h/g|>>\sqrt{2}+1$, $P_{s}$ and $P_{c}$ depend only on the SNR of the direct link and the reflecting link, respectively. This implies when the direct link is quite strong, the coupling effect between the primary and secondary transmissions in SR can be nearly ignored. 

\subsection{Scenario 3: Other values of $|h/g|$}
For other values of $|h/g|$, we give an approximation of the BER. According to~\cite{proakis2001digital}, the most probable errors occur when selecting the neighbor points as the true constellation point. Hence, an approximation of the BER is given by 
\begin{equation} \label{eq:BER-x-appro}
P_{x}\approx \frac{1}{|\mathcal{A}_{x}|}\sum_{i=1}^{|\mathcal{A}_{x}|}\sum_{x_{j}\in \mathcal{M}_{i}}P(\hat{x}=x_{j} \big| x = x_{i})\frac{d(x_{i}\rightarrow x_{j})}{\log_{2}(|\mathcal{A}_{x}|)},
\end{equation}
where $\mathcal{M}_{i}$ denotes the set of points that are adjacent to the point $x_{i}$.

In general, the decision regions are not regular, and thus it is difficult to calculate the integrals of $P(\hat{x}=x_{j} \big| x = x_{i})$ in \eqref{eq:conditional-probability}. It is shown that the pairwise error probability denoted by $P(x_{i}\rightarrow x_{j})$~\cite{proakis2001digital}, serves as an approximation for $P(\hat{x}=x_{j} \big| x = x_{i})$, given by
\begin{align} \label{eq:PEP}
P(\hat{x}=x_{j} \big| x = x_{i})&\approx P(x_{i}\rightarrow x_{j})\nonumber\\&
=P\left(|y-gx_{i}|^2> |y-gx_{j}|^2\right) \nonumber\\ &
\overset{a}{=}\mathcal{Q}\left(\mu |x_{i}-x_{j}|\right),
\end{align}
where $\mu=\frac{|g|}{\sqrt{2}\sigma}$, and `$a$' holds since $|g|^2|x_{i}-x_{j}|^2+2\Re\left\{z^{H}g(x_{i}-x_{j})\right\}$ follows the distribution of $\mathcal{N}(|g|^2|x_{i}-x_{j}|^2,2\sigma^{2}|g|^2|x_{i}-x_{j}|^2)$. It is reported in ~\cite{proakis2001digital} that this bound is quite tight for high SNR regimes.
Accordingly, the approximation of BER can be rewritten as 
\begin{align} \label{eq: BER-calculation}
P_{x}\approx\frac{1}{|\mathcal{A}_{x}|} \sum_{i=1}^{|\mathcal{A}_{x}|}\sum_{x_{j}\in \mathcal{M}_{i}}\mathcal{Q}\left(\mu|x_{i}-x_{j}|\right)\frac{d(x_{i}\rightarrow x_{j})}{\log_{2}(|\mathcal{A}_{x}|)}.
\end{align}


$P_{s}$ and $P_{c}$ can also be obtained using similar approximations as in \eqref{eq:BER-x-appro} and \eqref{eq:PEP}.
\section{Simulation Results} \label{sec-simulation}
\begin{table*} 	
	\caption{Optimal modulation scheme under different channel strength ratios}
	\begin{center}\label{table-1}
		\begin{tabular}{|c||c|c|c|c|c|}
			\hline
			$|h/g|$ & $0$ & $0.1$ & $1.5$ & $2.3$ & $4$\\
			\hline
			$\alpha^{\star}$ & $\cos(\pi/8)$ & $0.91$ & $0.18$ & $0$ & $0$ \\
			\hline
			$\beta^{\star}$ & -$\jmath\sin(\pi/8)$  & $-\jmath\times 0.42$ & $0.61-\jmath\times0.61$ & \makecell[c]{$|\beta^{\star}|=1$,
			$-0.38\pi\leq\theta^{\star}\leq-0.12\pi$ }& $|\beta^{\star}|=1$, random phase\\
			\hline
                {$K$} & {$\infty$ }& {$660$} & {$45$} & {$30$} & {$16$} \\
                \hline
		\end{tabular}
	\end{center}
\end{table*}
\begin{figure*}[t]
	\centering
	\begin{minipage}[t]{0.33\textwidth}
		\centering  
		\includegraphics[width=2.4in]{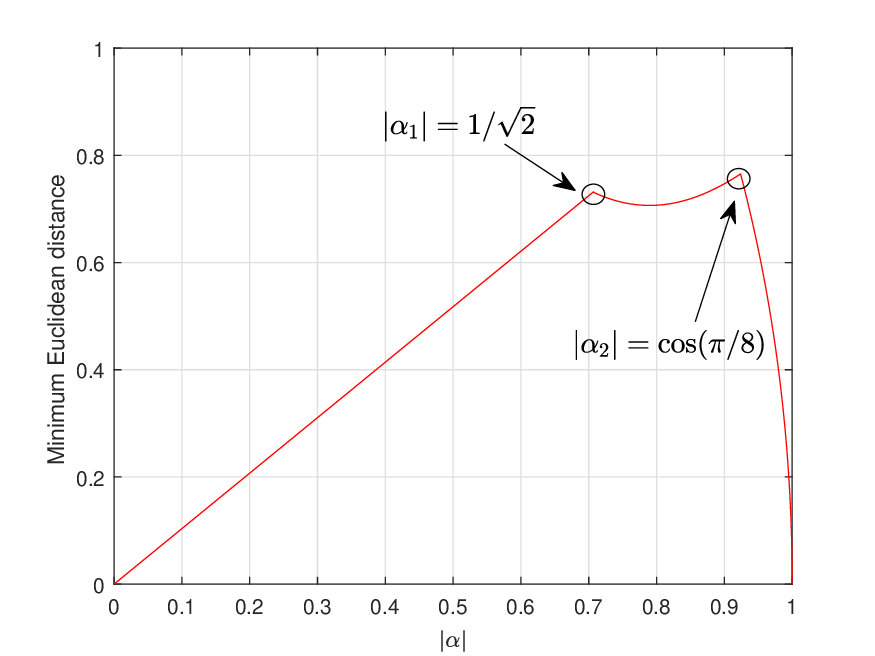} 
		\caption{The minimum Euclidean distance $D_{\min}^{*}(\alpha)$ versus $\alpha$, under $|h/g|=0$. }  
		\label{fig:fig_D_min}   
	\end{minipage}
	\begin{minipage}[t]{0.33\textwidth}
		\centering  
		\includegraphics[width=2.4in]{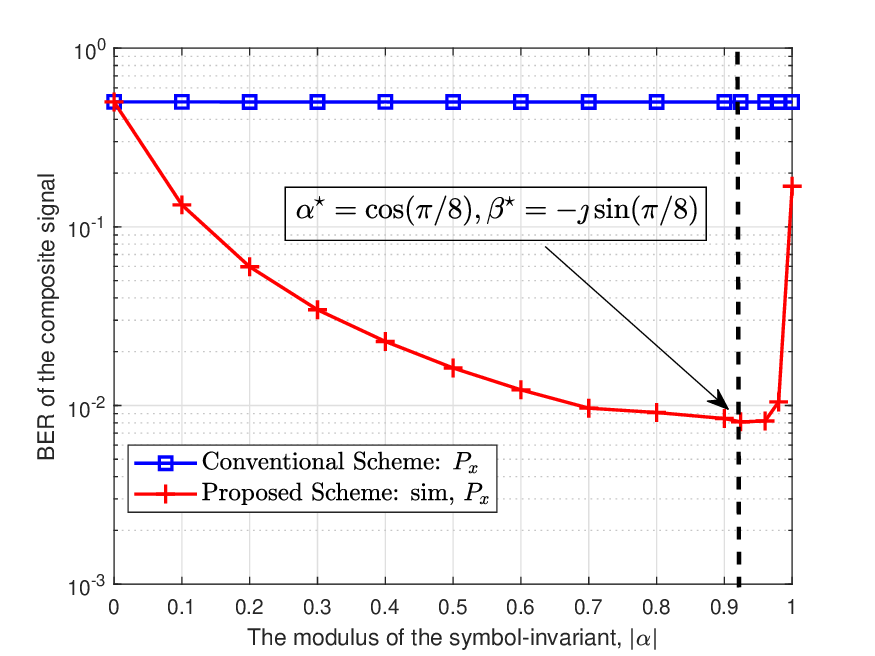} 
		\caption{BER of the composite signal (i.e., $P_{x}$) versus $\alpha$, under $\gamma_{b}=30$ dB.}  
		\label{fig:fig_8}   
	\end{minipage}
	\begin{minipage}[t]{0.32\textwidth}
		\centering  
		\includegraphics[width=2.4in]{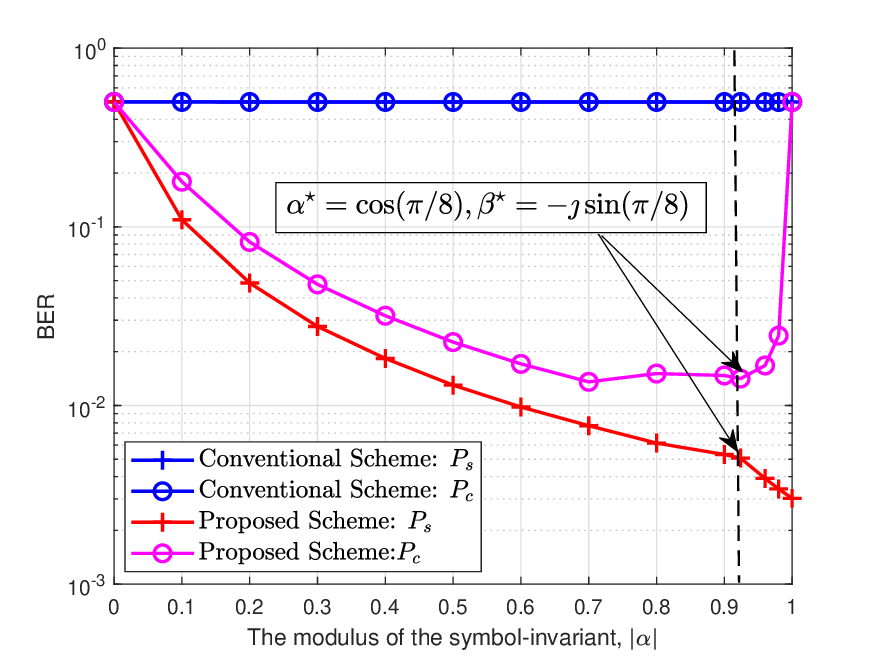}  
		\caption{BER of the primary and secondary signals (i.e., $P_{s}$ and $P_{c}$) versus $\alpha$, under $\gamma_{b}=30$ dB.}  
		\label{fig:fig_9} 
	\end{minipage}
\end{figure*}
In this section, simulation results are presented to validate the effectiveness of our proposed modulation scheme. We consider the conventional modulation scheme given by \eqref{eq:con} as a benchmark. 
Rayleigh fading is assumed for all involved channels. The large-scale path loss of the channel is modeled as $\mathrm{L}\left(d,\xi \right) = 10^{-3}d^{-\xi}$, where $d$ denotes the distance between two nodes, and $\xi$ denotes the path loss exponent~\cite{goldsmith2005wireless}.
We consider a topology where the coordinates of the PTx, the STx (RIS), and the C-Rx are set to $x_{T}=(0,0)$, $x_{B}=(75 \mathrm{m},10 \mathrm{m})$, $x_{R}=(80  \mathrm{m},0)$, respectively. Accordingly, the distances of the PTx--C-Rx link, the PTx-RIS link, and the RIS--C-Rx link can be calculated as $d_{\mathrm{T-R}}= \|x_{T}-x_{R}\|$, $d_{\mathrm{T-B}} = \|x_{T}-x_{B}\|$, $d_{\mathrm{B-R}} = \|x_{B}-x_{R}\|$, and the corresponding path loss exponents are set to $\xi_{T-R} = 3$, $\xi_{T-B}=2.1$, and $\xi_{B-R}=2.3$, respectively. In Table \ref{table-1}, we choose five representative channel strength ratios $|h/g|$ corresponding to the five cases discussed in Theorem \ref{thm: optimal-a-b}, and give the corresponding optimal design of $\alpha$ and $\beta$ as well as the required number of reflecting elements $K$. Note that the value of $|h/g|$ can be controlled by adjusting the number of RIS reflecting elements $K$ since $g=\sum_{k=1}^{K}|f_{k}||h_{r,k}|$ is dependent on $K$. 
Moreover, the reflecting link SNR is defined as $\gamma_{b}\triangleq\frac{p \sigma_{f}^{2}\sigma_{h,r}^{2}}{\sigma^{2}}$, where $\sigma_{f}^{2}=L(d_{T-B},\xi_{T-B})$ and $\sigma_{h,r}^{2}=L(d_{B-R},\xi_{B-R})$ denote the variance of each element in $\bm{g}$ and $\bm{h}_{r}$, respectively; the noise power is set to $\sigma^2=-100 \mathrm{dBm}$. By varying the transmit power $p$, we can get different values of $\gamma_{b}$.
Besides, all simulations are averaged over $10^6$ channel realizations. It should be noted that we assume instantaneous CSI is available to conduct simulation results, which can serve as an upper bound of system performance.

\subsection{Tradeoff Between the Performance of the Primary and Secondary Transmissions}
Taking $h=0$ as an example, we investigate the tradeoff between $P_{s}$ and $P_{c}$ under  $\gamma_{b}=30$ dB. In Fig. \ref{fig:fig_D_min}, we plot the minimum Euclidean distance $D_{\min}^{*}(\alpha)$ versus $\alpha$ by using the equation \eqref{eq: distance-alpha} under $|h/g|=0$. It can be seen that $D_{\min}^{*}(\alpha)$ increases over the interval $(0,\alpha_{1})$ and decreases over the interval $(\alpha_{2},1)$. Besides, it is observed that $D_{\min}^{*}(\alpha)$ first decreases  and then increases over the interval $(\alpha_{1},\alpha_{2})$. These phenomena validate our analysis in Sec. \ref{section: optimization-alpha}. Obviously, $D_{\min}^{*}(\alpha)$ attains its maximum at $\alpha_{2}=\cos(\pi/8)$, and thus we take $\alpha_{2}$ as the optimal $\alpha^{\star}$.

In Fig. \ref{fig:fig_8}, we plot the BER performance of the composite signal by varying $\alpha$ from $0$ to $1$, and the corresponding $\beta(\alpha)$ is calculated by using \eqref{eq: beta-alpha}.
Under the proposed scheme, it is shown that the $P_{x}$ first decreases when $\alpha$ varies from $0$ to $\alpha_{1}$, and increases when $\alpha$ varies from $\alpha_{2}$ to $1$. Moreover, we see that $P_{x}$  slightly increases and then decreases when $\alpha_{1}\leq \alpha \leq\alpha_{2}$. These phenomena correspond to the behaviour of  $D_{\min}^{*}(\alpha)$ versus $\alpha$, since the minimum Euclidean distance has a dominate impact on the BER performance. It is expected that $P_{x}$ is minimized when $\alpha^{\star}=\cos(\pi/8)$, $\beta^{\star}=-\jmath\sin(\pi/8)$, since $D_{\min}^{*}(\alpha)$ is maximized at this point.

In Fig. \ref{fig:fig_9}, under the same settings as Fig. \ref{fig:fig_8}, we plot the BER performance of the primary and secondary signals (i.e., $P_{s}$ and $P_{c}$) versus $\alpha$. We see that $P_{s}$ decrease with the increasing of $\alpha$. This means a larger symbol-invariant component will always lead to a better BER performance of the primary signal. Moreover, $P_{c}$ shows a similar behaviour as $P_{x}$ with respect to $\alpha$. This is because the Euclidean distance between the adjacent secondary signals, i.e.,  $2|\beta(\alpha)|$, given by \eqref{eq: beta-alpha} equals the minimum Euclidean distance $D_{\min}^{*}(\alpha)$ given by \eqref{eq: distance-alpha}. 
Besides, we see that decreasing the energy for the symbol-varying component $|\beta|$ will not decrease the BER performance of the secondary signal.
This is because although the energy for $|\beta|$ is reduced, more energy for the symbol-invariant component $\alpha$ can be used to assist the primary transmission. This in turn helps improve the performance of the secondary transmission thanks to the coupling effect.
Considering both $P_{s}$ and $P_{c}$, the performance of the primary and secondary signals can strike a balance at the point $\alpha^{\star}=\cos(\pi/8)$, $\beta^{\star}=-\jmath\sin(\pi/8)$.

\subsection{BER Performance}
\begin{figure}[!t]
	\centering 
	\setlength{\abovecaptionskip}{-0.05cm}
	\setlength{\belowcaptionskip}{-0.05cm}
	\subfigure[ $|h/g|=0$,  $|h/g|=0.1$, $|h/g|=1.5$] {
		\label{fig:fig2}
		\includegraphics[width=0.42\textwidth]{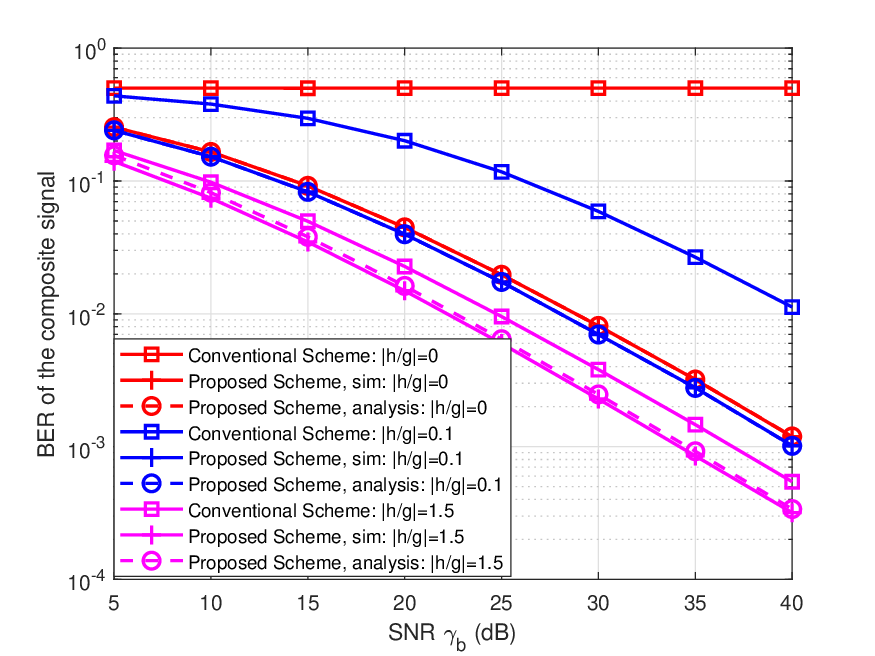}
	}
	\subfigure[$|h/g|=2.3$, $|h/g|=4$, $|h/g|=20$]{
		\label{fig:fig3}
		\includegraphics[width=0.42\textwidth]{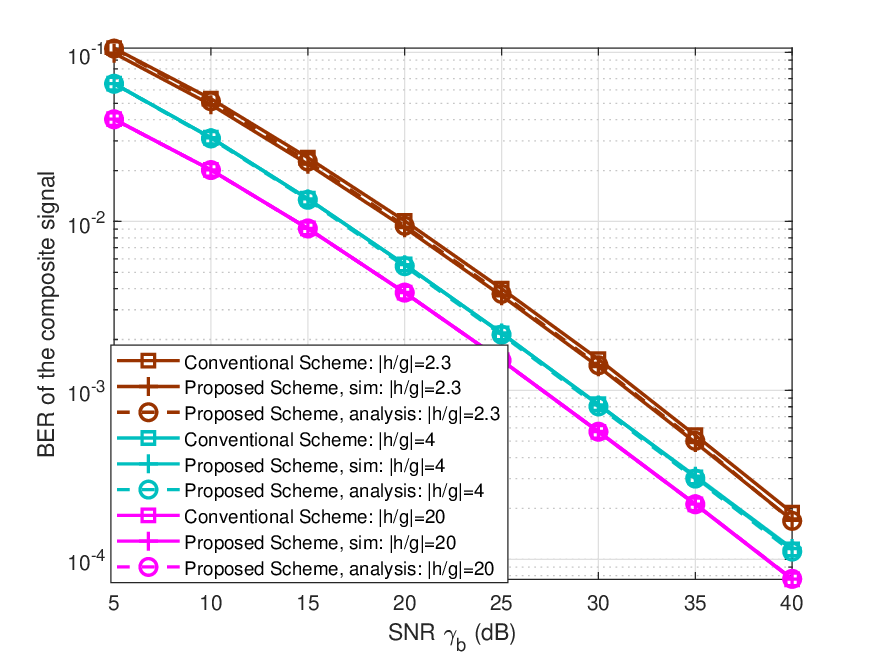}
	}
	\caption{ BER of the composite signal $x$ (i.e., $P_{x}$) versus the reflecting link SNR $\gamma_{b}$.}\label{figall-2}
\end{figure}

Fig. \ref{figall-2} shows the BER of the composite signal $x$ versus the reflecting link SNR $\gamma_{b}$, under different channel strength ratios. 
One can observe that the analytical results are almost identical to the simulations, which corroborates our BER analysis.
From Fig. \ref{figall-2} (a), we see that if the conventional modulation scheme is applied and  $|h/g|=0$, the C-Rx cannot decode both the primary and secondary signals due to the ambiguity problem defined in Sec. \ref{sec-system-model}. 
Accordingly, the BER of $x$ is $0.5$ regardless of the reflecting link SNR. However, such an ambiguity problem no longer exists when applying our proposed modulation scheme. This is because the introduction of the symbol-invariant component (i.e., $\bm{\Phi}_{1}=\alpha \bm{\Phi}$) can be viewed as a virtual direct link.
Moreover, one can see that the proposed modulation scheme outperforms the conventional counterpart, especially when the direct link is relatively weak. For the case of $|h/g|=0.1$, the proposed scheme achieves an SNR gain of $12$ dB over the conventional scheme when $P_{x}=10^{-2}$.
From Fig. \ref{fig:fig2} and \ref{fig:fig3}, we observe that the performance gap between the conventional modulation scheme and the proposed modulation scheme becomes smaller as the channel strength ratio becomes larger. In particular, when $|h/g|=4$ and $|h/g|=20$, the proposed design shown in Table \ref{table-1} is equivalent to the conventional design, therefore leading to the same BER performance.
\begin{figure}[!t]
	\centering
	\setlength{\abovecaptionskip}{-0.05cm}
	\setlength{\belowcaptionskip}{-0.05cm}
	\subfigure[$|h/g|=0$,  $|h/g|=0.1$, $|h/g|=1.5$] {
		\label{fig:fig4}
		\includegraphics[width=0.42\textwidth]{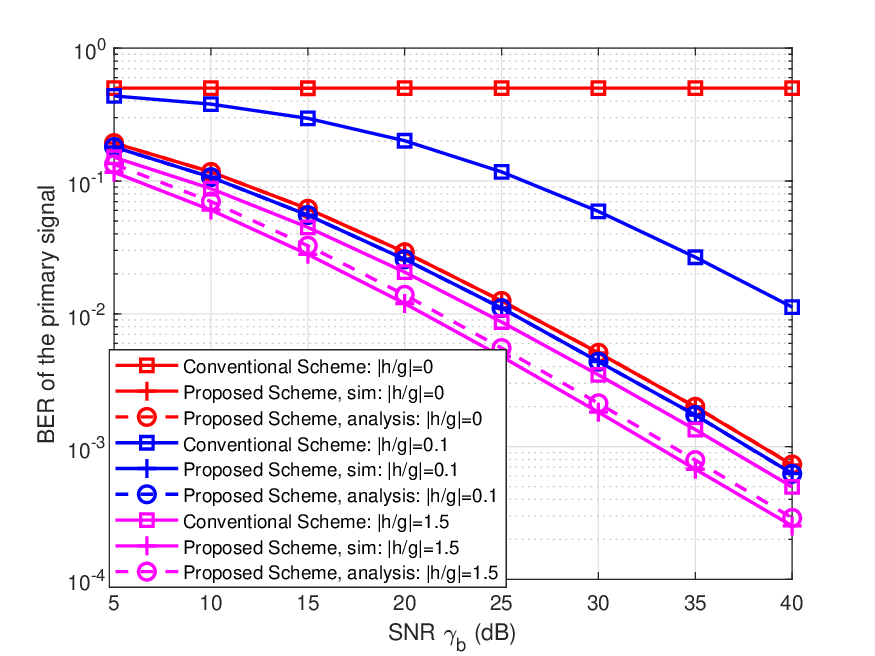}
	}
	\subfigure[$|h/g|=2.3$,  $|h/g|=4$, $|h/g|=20$]{
		\label{fig:fig5}
		\includegraphics[width=0.42\textwidth]{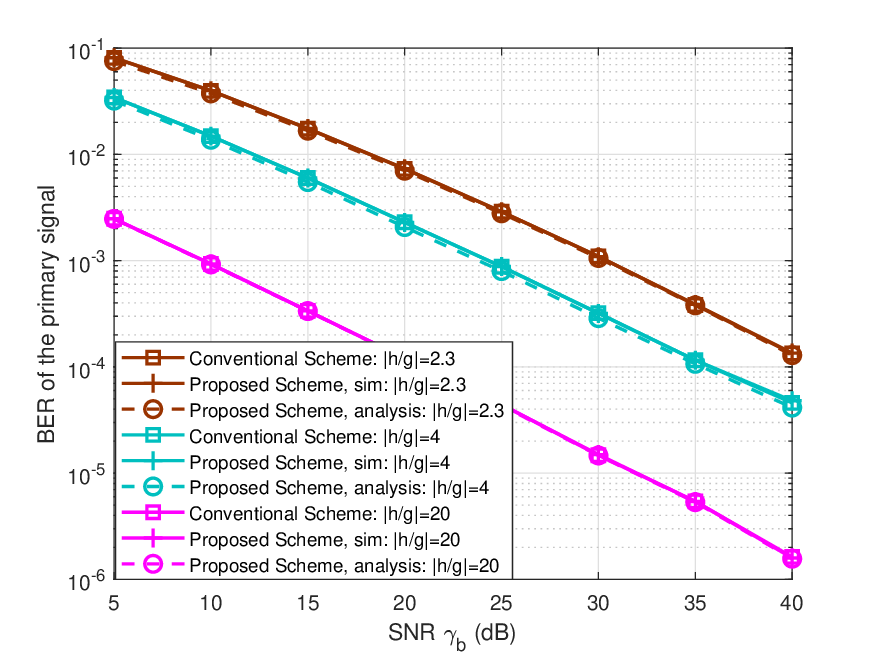}
	}
	\caption{ BER of the primary signal $s$ (i.e., $P_{s}$) versus the reflecting link SNR $\gamma_{b}$.}\label{figall-3}
\end{figure}

Next, we show the impact of the proposed design on the performance of the primary signal. Fig. \ref{figall-3} illustrates $P_{s}$ versus the reflecting link SNR $\gamma_{b}$ under different channel strength ratios. From Fig. \ref{fig:fig4}, for the case of $|h/g|=0$, the C-Rx cannot recover both the primary and secondary signals due to the ambiguity problem. By adopting the proposed design, $P_{s}$ decreases with $\gamma_{b}$. The reason is that when the direct link is weak, we introduce a symbol-invariant component at the RIS to assist the primary transmission.
For the case of $|h/g|=0.1$, the proposed design outperforms the conventional design, which achieves an SNR gain of $15$ dB when $P_{s}=10^{-2}$.
From Fig. \ref{fig:fig5}, as the channel strength ratio increases, the BER performance of conventional design approaches that of the proposed modulation design. This is because when the direct link becomes stronger, the symbol-invariant component related to $\alpha$ used for enhancing the primary transmission will correspondingly decrease. Especially for the case of $|h/g|>\sqrt{2}+1$ defined in Theorem \ref{thm: optimal-a-b}, $P_{s}$ is dominated by the direct link, and thus the primary transmission
does not need the symbol-invariant component anymore. In this case, the proposed design degrades to the conventional design. Moreover, when $P_{s}=10^{-3}$, we increase $|h/g|$ from $4$ to $20$ will introduce an SNR gain at $14$ dB, due to $10\log_{10}(\frac{20}{4})^{2}\approx14$. Besides, when $|h/g|=20$, the BER performance of primary transmission will approach $\mathcal{Q}(|h|/\sigma)$.

\begin{figure}[!t]
	\centering
	\setlength{\abovecaptionskip}{-0.05cm}
	\setlength{\belowcaptionskip}{-0.05cm}
	\subfigure[$|h/g|=0$,  $|h/g|=0.1$, $|h/g|=1.5$] {
		\label{fig:fig6}
		\includegraphics[width=0.42\textwidth]{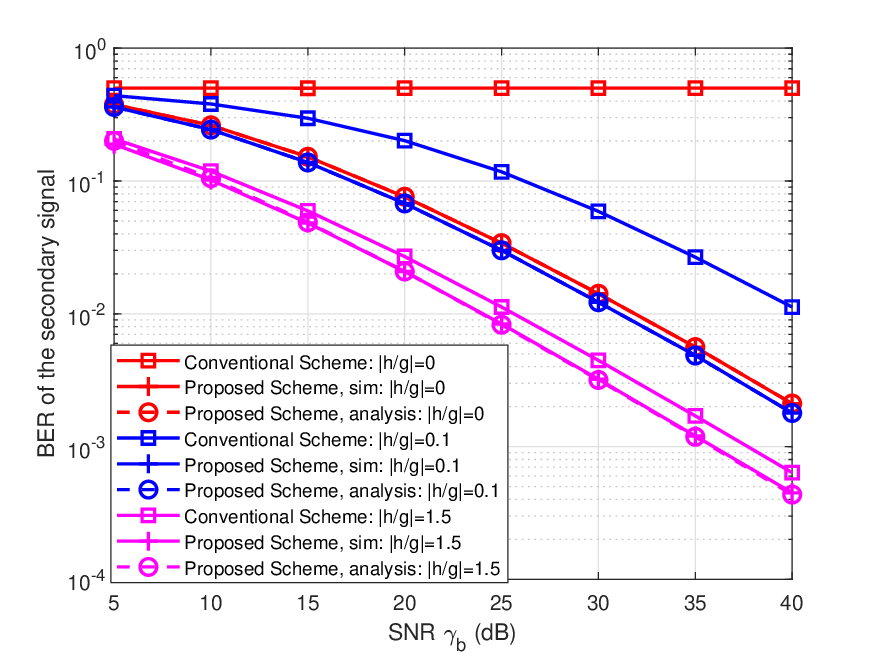}
	}
	\subfigure[$|h/g|=2.3$,  $|h/g|=4$, $|h/g|=20$]{
		\label{fig:fig7}
		\includegraphics[width=0.42\textwidth]{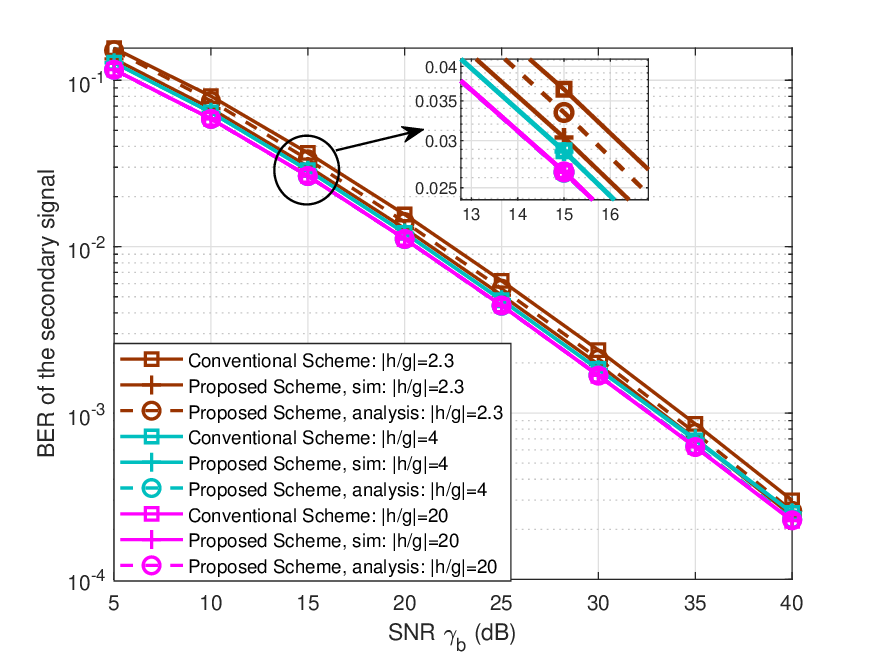}
	}
	\caption{ BER of the secondary signal $c$ (i.e., $P_{c}$) versus the reflecting link SNR $\gamma_{b}$.}\label{figall-4}
\end{figure}
Finally, we show the impact of the proposed design on the secondary transmission. Fig. \ref{figall-4} illustrates $P_{c}$ versus the reflecting link $\gamma_{b}$ under different channel strength ratios. 
Similarly, when $|h/g|=0$, we have $P_{c}=0.5$ under the conventional modulation scheme.
For the case of $|h/g|=0.1$, compared with the conventional scheme, the proposed scheme achieves an SNR gain at almost $9$ dB when $P_{c}=10^{-2}$.
From Fig. \ref{fig:fig7}, we observe a similar phenomenon that the performance gap between the proposed scheme and the conventional scheme becomes negligible when $|h/g|$ goes larger. Besides, we see that when we increase $|h/g|$ from $4$ to $20$, $P_{c}$ only decreases marginally and approaches $\mathcal{Q}(\sqrt{2}|g|/\sigma)$. This is because the lower bound of $P_{c}$ is $\mathcal{Q}(\sqrt{2}|g|/\sigma)$ with known $s$. When $|h/g|$ is very large, $s$ will be decoded almost perfectly, and $P_{c}$ will approach this lower bound.

\subsection{Impact of CSI Errors and Structural Mode Reflection} \label{sec;structural}
\begin{figure}[t]
	\centering
	\setlength{\abovecaptionskip}{-0.05cm}
	\setlength{\belowcaptionskip}{-0.05cm}
	\begin{minipage}[t]{0.49\textwidth}
		\centering  
		\includegraphics[width=3in]{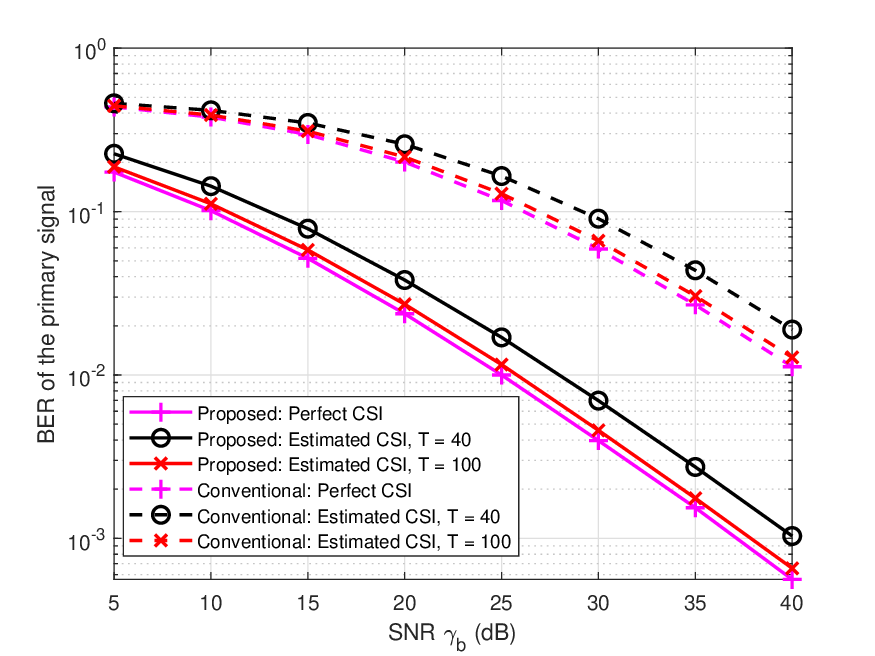} 
		\caption{BER of the primary signal versus reflecting link SNR under different channel estimation errors. }  
		\label{fig:Imperfect_primary}   
	\end{minipage}
	\begin{minipage}[t]{0.49\textwidth}
		\centering  
		\includegraphics[width=3in]{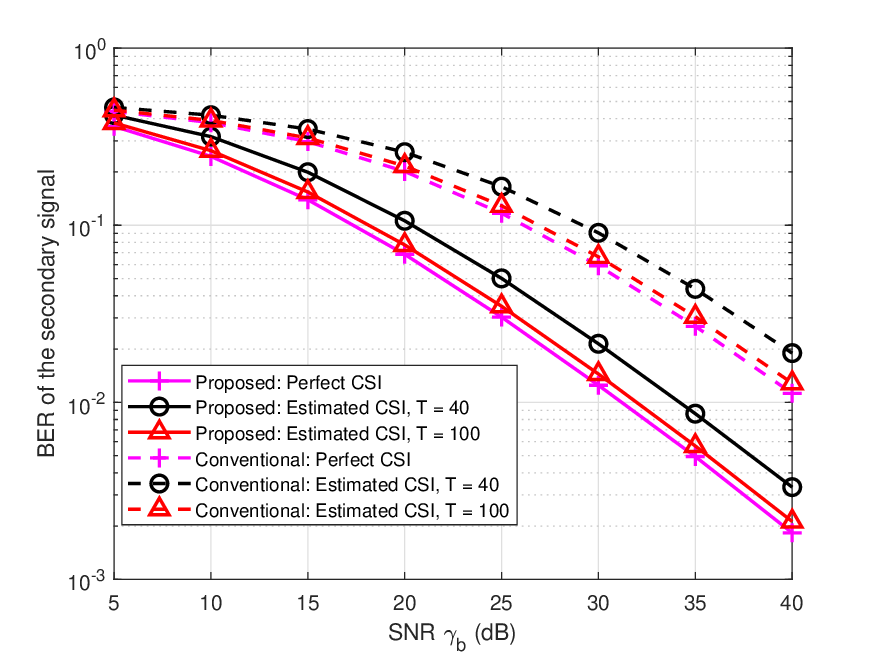} 
		\caption{BER of the secondary signal versus reflecting link SNR under different channel estimation errors}  
		\label{fig:Imperfect_secondary}   
	\end{minipage}
\end{figure}
In this subsection, we first study the impact of CSI errors on our proposed modulation design. Here, we are interested in estimating the direct channel $h$ and the cascaded channel $\bm{v}=\mathrm{diag}(\bm{h}_{r})\bm{f}$ within $T$ training slots, where the received signal can be written in a vector form, given by
\begin{align} \label{eq: received-error}
  \bm{y}=\sqrt{p}\bm{S}\bm{\Phi}_{0}\bm{h}_{\mathrm{est}}+\bm{z}.
\end{align}

In \eqref{eq: received-error}, $\bm{S}\in\mathbb{C}^{K\times K}$ is a diagonal matrix denoting the pilot symbols sent by the PTx; $\bm{\Phi}_{0}\in\mathbb{C}^{T\times (K+1)}$ denotes the training reflecting patterns of RIS, where $t$-th row is given by $[1, \bm{\phi}_{t}]\in \mathbb{C}^{1\times (K+1)}$; $\bm{h}_{\mathrm{est}}=[h;\bm{v}]$ denotes the channels to be estimated.

Following prior work~\cite{swindlehurst2022channel}, the least-square (LS) estimator can be utilized here\footnote{The estimation overhead is related to the number of pilot symbols, which should be larger than the number of reflecting elements for the least square method. Some advanced methods designed to reduce the estimation overhead can be found in \cite{zheng2022survey}, e.g., element-grouping and matrix factorization.}, where the variance of the estimation error, namely MSE, can be derived as
\begin{align} \nonumber
    \mathrm{MSE}=\epsilon  = \bm{\mathbb{E}} \left[
    \|\bm{h}_{\mathrm{est}}-\hat{\bm{h}}_{\mathrm{est}}\|^2
    \right]\nonumber  =\mathrm{tr} \left(\frac{\sigma^2}{p}
    (\bm{\Phi}_{0}^{H}\bm{S}^{H} \bm{S}\bm{\Phi}_{0})^{-1}\right).
\end{align}
For simulations, we use the DFT codebook as the training reflecting patterns for RIS and the pilot symbols satisfy unit-modulus constraint $\bm{S}^{H}\bm{S}=\bm{I}_{T\times T}$.
\begin{figure}[t]
	\centering
	\setlength{\abovecaptionskip}{-0.05cm}
	\setlength{\belowcaptionskip}{-0.05cm}
	\begin{minipage}[t]{0.49\textwidth}
		\centering  
		\includegraphics[width=3in]{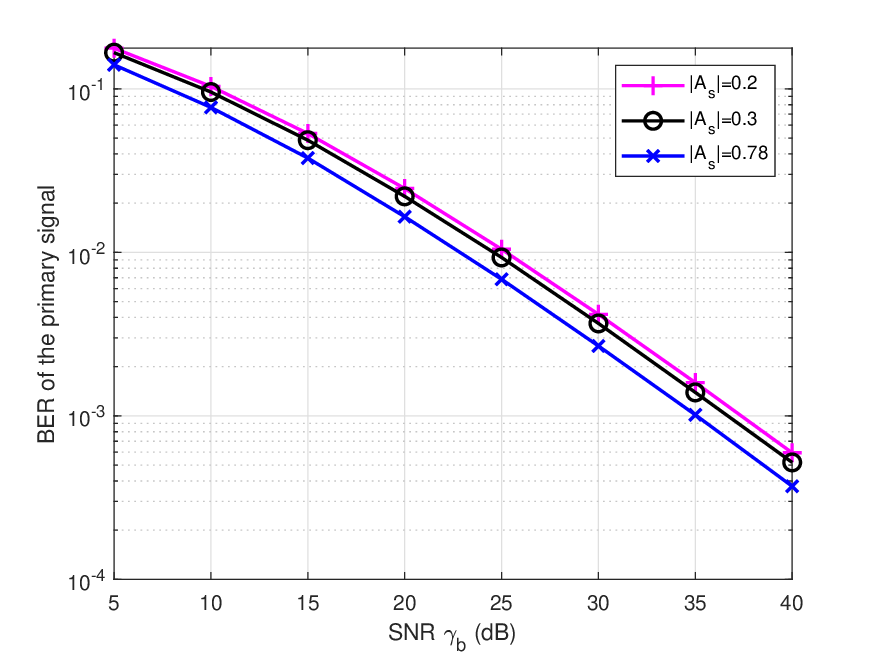} 
		\caption{BER of the primary signal versus reflecting link SNR under different structural mode reflection. }  
		\label{fig:Structural_primary}   
	\end{minipage}
	\begin{minipage}[t]{0.49\textwidth}
		\centering  
		\includegraphics[width=3in]{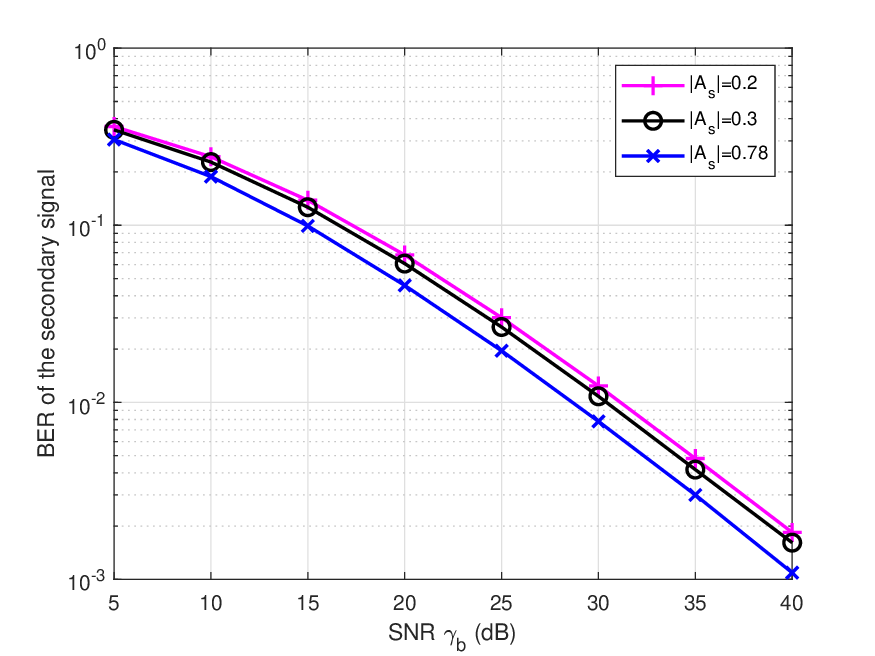} 
		\caption{BER of the secondary signal versus reflecting link SNR under different structural mode reflection.}  
		\label{fig:Structural_secondary}   
	\end{minipage}
\end{figure}

In Fig. \ref{fig:Imperfect_primary} and Fig. \ref{fig:Imperfect_secondary}, we plot the BERs of the primary and secondary signals under estimated CSI with different numbers of training slots $T$, under $|h/g|=0.1$. First, it can be observed that with the increase of training slots, a better BER performance with estimated CSI can be achieved. This is obvious since increasing $T$ can decrease the channel estimation errors. When $T$ is large enough, the BER performance of the estimated CSI can approach that of the perfect CSI. Moreover, it can be seen that even with the estimated CSI, our proposed method still outperforms the conventional scheme, which validates the effectiveness of our proposed modulation scheme.

Then, the impact of structural mode reflection on the BER performance is studied. Following \cite{kimionis2014increased}, we choose three values of $\bm{A}_{s}$ for simulations, which are $\bm{A}_{s}=(0.6047+\jmath0.5042)\bm{I}_{K\times K}$, $\bm{A}_{s}=(0.2954-\jmath0.0524)\bm{I}_{K\times K}$, and $\bm{A}_{s}=(0.1593-\jmath0.1209)\bm{I}_{K\times K}$.
In this case, since the value of $h$ is related to the value of 
$\bm{A}_s$, $|h/g|$ can be calculated according to different $\bm{A}_s$ based on \eqref{eq-direct-link} and \eqref{eq:reflecting-link}.
From Fig. \ref{fig:Structural_primary} and Fig. \ref{fig:Structural_secondary}, we can see that with the increase of the modulus of structural mode reflection, $|\bm{A}_{s}|$, the BER will decrease for both the primary and secondary transmissions. This is because the increase of $|A_{s}|$ can help enhance the equivalent direct link, which could in turn enhance the BER performance.
Such a phenomenon suggests that it is important to take care of the structural mode reflection $\bm{A}_{s}$, in addition to the antenna mode reflection of RIS.
\subsection{Impact of Spatial Correlation of RIS-related Channels}
In practice, the RIS-related channels are spatially correlated, which is investigated in \cite{bjornson2020rayleigh}. Assume a uniform planar array deployed at the RIS with $K_H$ and $K_V$ reflecting elements per row and per column, respectively (i.e., $K=K_{H}K_{V}$).
Based on the results of \cite{bjornson2020rayleigh}, we can model the channel correlation matrix as $\bm{R}\in\mathbb{C}^{K\times K}$, whose $(i,j)$-th element is given by
\begin{align} \label{eq-correlation}
    \bm{R}_{i,j}= \mathrm{sinc}(\frac{2\Vert\bm{u}_{i}-\bm{u}_{j}\Vert}{\lambda}).
\end{align}
In \eqref{eq-correlation}, $\mathrm{sinc(x)}=\frac{\sin(\pi x)}{\pi x}$ denotes the $\mathrm{sinc}$ function; $\lambda$ denotes the wavelength of the carrier signal; $\bm{u}_{i}= (H(i)d_{H},V(i)d_{V})$, where $d_H$ and $d_V$ denotes the element spacing in the horizontal and vertical directions, and generally we have $d_H=d_V=d$ called element spacing;
$H(i)=\mathrm{mod}(i-1,K_{H})$ and $V(i)=\lfloor(i-1)/K_{H}\rfloor$. Here, $\mathrm{mod}(\cdot)$ and $\lfloor\cdot\rfloor$ denote the modulus operation and integer floor operation, respectively. Also, $\bm{u}_{j}$ can be written similarly. 

\begin{figure}  
	\centering  
	\setlength{\abovecaptionskip}{0.cm}
	\includegraphics[width=3in]{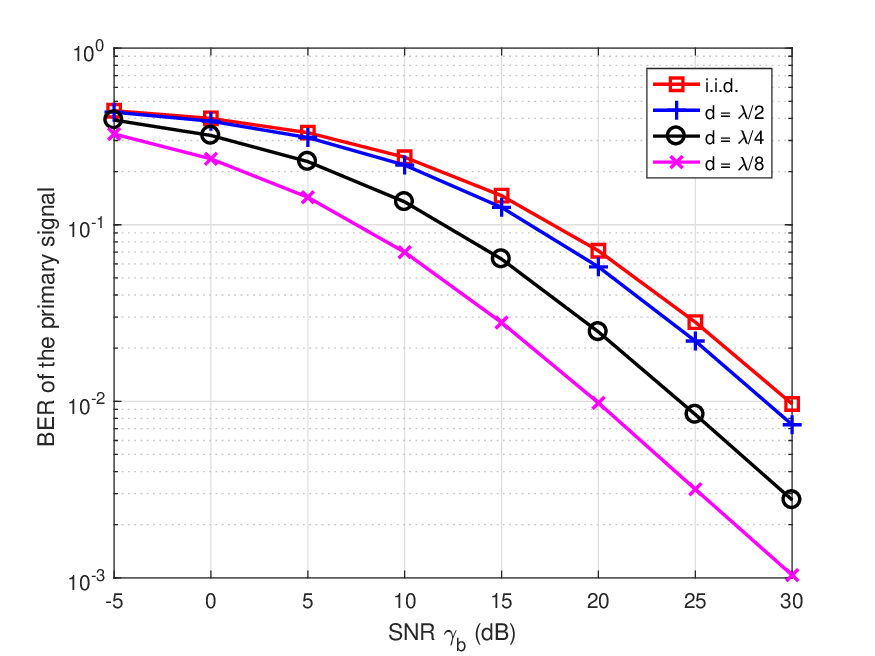}  
	\caption{BER of the primary signal versus reflecting link SNR under different element spacing of RIS.}  
	\label{fig:channel-correlation}  
\end{figure}

By collecting all the elements $\bm{R}_{i,j}$ via \eqref{eq-correlation}, we can obtain the channel correlation matrix $\bm{R}$. Then, PT--RIS and RIS--C-Rx channels can be rewritten as $\bm{f}=L_{1}\bm{R}^{\frac{1}{2}}\tilde{\bm{f}}$ and $\bm{h}_{r}=L_{2}\bm{R}^{\frac{1}{2}}\tilde{\bm{h}_{r}}$, where $L_{1}$ and $L_{2}$ denote their path loss coefficients, respectively; $\bm{R}^{\frac{1}{2}}$ denotes the square root of the spatial correlation matrix $\bm{R}$, $\tilde{\bm{f}}\sim \mathcal{CN}(0,1)$ and $\tilde{\bm{h}_{r}}\sim \mathcal{CN}(0,1)$ account for complex Gaussian random vectors with i.i.d entries. Then, when we obtain the spatially correlated channels $\bm{f}$ and $\bm{h}_{r}$, our proposed methodology can be directly applied and obtain corresponding optimization parameters $\alpha$, $\beta$, and RIS phase shifts. Based on the above process, in Fig. \ref{fig:channel-correlation}, we plot the BER of the primary signal versus the reflecting link SNR under different element spacing of RIS with $|h/g|=0$. It is shown that the BER performance becomes better when the element spacing gets smaller. This coincides with the result shown in \cite{ge2022ris}, and it is as expected since the element spacing becomes smaller, the beamforming gain becomes larger accordingly.

\section{Conclusion} \label{sec-con}
In this paper, we have investigated the fundamental modulation design for RIS-assisted SR. First, we have proposed a novel modulation scheme that divides the phase-shift matrix into two components, i.e., the symbol-invariant and symbol-varying components.
The former is used to assist the primary system and the latter is used to carry the secondary signal.
Then, we have studied the problem of optimizing these two components to minimize the BER of the composite signal formed by the primary and secondary signals. After that, we have presented the optimal solutions using geometrical analysis and have analyzed the theoretical BER performance. Simulation results have shown that our proposed modulation scheme could strike a balance between the BER performance of the primary and secondary transmissions.
\begin{appendices} 

        \section{} \label{Appendix-prop1}
         Proposition \ref{PropWeightedChannel} can be proved by exploiting the structure of the Euclidean distance terms of the composite signal. Specifically, the Euclidean distance between the composite signals $x_{m}$ and $x_{l}$ can be written as 
        \begin{align}
            D_{m,l}^{2}\!=\!\frac{1}{|g|^2}|(h\!+\!\bm{h}_{r}^{H}\bm{\Phi}_{1}\bm{f})(s_{m}\!-\!s_{l})            \!+\!\bm{h}_{r}^{H}\bm{\Phi}_{2}\bm{f}(s_{m}c_{m}\!-\!s_{l}c_{l})|^2. \nonumber
        \end{align}
        
        The Euclidean distances can be categorized into the following three types.
        \begin{itemize}
            \item \emph{Case 1:} $s_{m}=s_{l}$, $c_{m}\neq c_{l}$. In this case, $D_{m,l}^{2}$ can be expressed as
            \begin{align}
                D_{m,l}^{2} = \frac{1}{|g|^2}|\bm{h}_{r}^{H}\bm{\Phi}_{2}\bm{f}|^2|c_{m}-c_{l}|^2.
            \end{align}
            If we want to maximize this distance term, it is obvious that the phase shifts $\bm{\Phi}_{2}$ should be designed to align the reflecting link, given by  $\bm{\Phi}_{2}\!=\!\mathrm{diag}[e^{- \jmath (\phi_{0}+\angle (f_{k}h_{r,k}))},\cdots,e^{- \jmath (\phi_{0}+\angle (f_{K}h_{r,K}))}]$, where $\phi_{0}$ denotes an arbitrary common phase shift.
            \item  \emph{Case 2:} $s_{m}\neq s_{l}$, $c_{m}\neq c_{l}$, and $s_{m}c_{m}=s_{l}c_{l}$. In this case, we have
            \begin{align}
                D_{m,l}^{2} = \frac{1}{|g|^2}|h+\bm{h}_{r}^{H}\bm{\Phi}_{1}\bm{f}|^2|s_{m}-s_{l}|^2,
            \end{align}
         In this case, the Euclidean distance can be maximized by designing $\bm{\Phi}_{1}$ in a way to align the reflecting link with the direct link, given by  $\bm{\Phi}_{1}\!=\!\mathrm{diag}[e^{ \jmath (\angle(h)-\angle (f_{k}h_{r,k}))},\cdots,e^{\jmath (\angle(h)-\angle (f_{K}h_{r,K}))}]$.
        \item \emph{Case 3:} For the remaining relationships between $\{s_{m},c_{m}\}$ and $\{s_{l},c_{l}\}$, $D_{m,l}^{2}$ can be written as
        \begin{align} \nonumber
            &D_{m,l}^{2} =\frac{1}{|g|^2}( |h+\bm{h}_{r}^{H}\bm{\Phi}_{1}\bm{f}|^2|s_{m}-s_{l}|^2+|\bm{h}_{r}^{H}\bm{\Phi}_{2}\bm{f}|^2
            \\
            & \quad  |s_{m}c_{m}-s_{l}c_{l}|^2 +2\Re\{(h+\bm{h}_{r}^{H}\bm{\Phi}_{1}\bm{f})^{H}\bm{h}_{r}^{H}\bm{\Phi}_{2}\bm{f} 
           \nonumber \\
            & \quad 
            (s_{m}-s_{l})^{H}(s_{m}c_{m}-s_{l}c_{l})\}).
        \end{align}
        In this case, the Euclidean distance is related to both $\bm{\Phi}_{1}$ and $\bm{\Phi}_{2}$. Fortunately, it can be observed that the first two terms of $D_{m,l}^{2}$ in \emph{Case 3} correspond to the Euclidean distances in \emph{Case 1} and \emph{Case 2}, respectively. Therefore, the design of $\bm{\Phi}_{1}$ and $\bm{\Phi}_{2}$ in \emph{Case 1} and \emph{Case 2} can be directly applied here, and the common phase shift $\phi_{0}$ in $\bm{\Phi}_{2}$ can be determined by maximizing the third term in \emph{Case 3} according to different $\{s_{m},c_{m}\}$ and $\{s_{l},c_{l}\}$, $D_{m,l}^{2}$.
        \end{itemize}
        
        In summary, we find the similarity between $\bm{\Phi}_{1}$ and $\bm{\Phi}_{2}$, whose common idea is to align the reflecting link. However, due to the modulus constraint of RIS phase shifts \eqref{eq:modulus-constraint-1} and \eqref{eq:modulus-constraint-2}, the amplitude of $\bm{\Phi}_{1}$ and $\bm{\Phi}_{2}$ cannot be unity. To guarantee the above phase alignment function of $\bm{\Phi}_{1}$ and $\bm{\Phi}_{2}$,  $\bm{\Phi}_{1}+c\bm{\Phi}_{2}$ can be equivalently written as the following form
        \begin{align}
            \bm{\Phi}_{1}+c \bm{\Phi}_{2}=\bm{\Phi} (\alpha+\beta c),
        \end{align}
        where $\bm{\Phi}$ is the common phase-shift matrix shared by $\bm{\Phi}_{1}$ and $\bm{\Phi}_{2}$, given by $\bm{\Phi}=e^{\jmath \angle (h)}\mathrm{diag}\left[e^{- \jmath \angle (f_{k}h_{r,k})},\cdots,e^{- \jmath \angle (f_{K}h_{r,K})}\right]$;
        $\alpha\in \mathbb{R}$ and $\beta\in \mathbb{C}$ denote two weighted parameters.
         Through this transformation, we can control $\alpha$ and $\beta$ to satisfy the modulus constraints. 
	\section{} \label{Appdendix: rotate}
\begin{figure}[t]
		\centering
		\captionsetup{font={scriptsize}}
		\begin{minipage}[t]{0.49\linewidth}
			\centering  
                \includegraphics[width=0.9\linewidth]{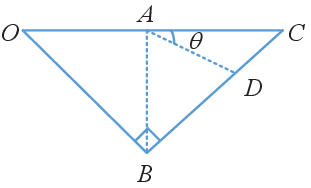} 
			\caption{Geometrical sketch of constraints $\mathrm{C1}$ and $\mathrm{C2}$.}  
			\label{fig:fig_AppendixA}   
		\end{minipage}
		\begin{minipage}[t]{0.49\linewidth}
			\centering  
			\includegraphics[width=0.9\linewidth]{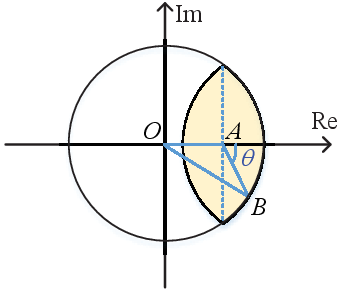}  
			\caption{Geometrical sketch of constraint $\mathrm{C3}$ and $\mathrm{C4}$.}  
			\label{fig:fig_AppendixB} 
		\end{minipage}
	\end{figure}

	 Since we assume $\angle\alpha=\angle (h/g)=0$, according to the constraints $\mathrm{C1,C2}$, i.e.,  $\angle\left(\frac{h}{g}+\alpha+\beta\right)\leq0$ and $\angle\left(\frac{h}{g}+\alpha-\beta\right)\geq0$, we have the phase of $\beta$, i.e., $\theta\in(-\pi,0)$. As illustrated in Fig. \ref{fig:fig_AppendixA}, line $O-A$ denotes the vector $\frac{h}{g}+\alpha$. The constraints in $\mathrm{C1}$ and $\mathrm{C2}$ form the feasible region of $\beta$, which is shown as the isosceles right triangle $O-B-C$.
	 Specifically, $\beta$ is the vector beginning with node $A$, and ending with the arbitrary node in the triangle $O-B-C$. 
	  Denote the arbitrary node on line $B-C$ by $D$. We are interested in the triangle $A-D-C$, where the angle $\angle A-C-D=\frac{1}{4}\pi$, angle $\angle D-A-C =-\theta\geq0$. Then the angle
	 $\angle A-D-C$ can be calculated as $\angle A-D-C= \frac{3}{4}\pi+\theta$. In terms of the law of sines, we have the following equation
	 \begin{equation}
	 \frac{|AD|}{\sin(\frac{1}{4}\pi)}=\frac{|AC|}{\sin(\frac{3}{4}\pi+\theta)},
	 \end{equation}
	where $|AC| =\left|\frac{h}{g}+\widetilde{\alpha} \right|$, $|AD|$ denotes the length of $\beta$.
	Accordingly, the length of vector $A-D$ is calculated as $|AD|=\frac{1}{\cos(\theta)-\sin(\theta)}\left|\frac{h}{g}+\widetilde{\alpha} \right|$.  Denote the arbitrary node on line $O-B$ by $E$, and let $\beta$ be the vector $A-E$, then we have $|AE|=|\beta|=\frac{1}{-\cos(\theta)-\sin(\theta)}\left|\frac{h}{g}+\widetilde{\alpha}\right|$ due to symmetry.
	\section{}\label{Appendex_modulous-one}
 	Given $\widetilde{\alpha}$, the constraints $\mathrm{C3,C4}$, i.e., $|\widetilde{\alpha}+\beta|\leq1$ and $|\widetilde{\alpha}-\beta|\leq1$, define the feasible region of $\beta$, which is shown as the yellow region in Fig. \ref{fig:fig_AppendixB}. Denote the arbitrary node on the boundary by $B$, the nodes $O$, $A$, and $B$ constitutes a triangle $O-A-B$, where the length of side $O-A$ is denoted by $|OA|=\widetilde{\alpha}$, the length of side $O-B$ is denoted by $|OB|=1$, and the internal angle contained between sides $O-A$ and $O-B$ is denoted by $\angle ~A-O-B =\pi+\theta$ since $\theta\leq0$. According to the law of cosines, we have 
	\begin{equation} \label{eq:ZX}
	|OB|^2= |OA|^2+|AB|^2-2|OA||AB|\cos(\pi+\theta).
	\end{equation}
	
	From \eqref{eq:ZX}, the $|AB|$ can be calculated as $|AB|=\sqrt{1-|\widetilde{\alpha}|^2\sin^2(\theta)}-|\widetilde{\alpha}|\cos(\theta)$. Due to symmetry, the boundary of the feasible region of $\beta$ can be expressed as $\sqrt{1-|\widetilde{\alpha}|^2\sin^2(\theta)}-|\widetilde{\alpha}|\cos(\theta)(-1)^{\mathbb{I}_{2}(\theta)}$, where $\mathbb{I}_{2}(\theta)$ is a indicator function defined in \eqref{eq: c4-ba}.
	\section{} \label{Appendix-D-min-general}
	All the distances can be obtained through simple calculation, given by the following set $\mathcal{D}=\{2|\beta|, \sqrt{2}|\frac{h}{g}+\widetilde{\alpha}-\beta|,\sqrt{2}|\frac{h}{g}+\widetilde{\alpha}-\jmath\beta|,\sqrt{2}|\frac{h}{g}+\widetilde{\alpha}+\beta|,
2|\frac{h}{g}+\widetilde{\alpha}|, \sqrt{2}|\frac{h}{g}+\widetilde{\alpha}+\jmath\beta|,2|\frac{h}{g}+\widetilde{\alpha}-\beta|,2|\frac{h}{g}+\widetilde{\alpha}+\beta|\}$.
Obviously, $2|\frac{h}{g}+\widetilde{\alpha}+\beta|$ and $2|\frac{h}{g}+\widetilde{\alpha}-\beta|$ will not be the minimum of the set $\mathcal{D}$.
One can observe that given $\widetilde{\alpha}$ and $|\beta|$, the distance terms $2|\frac{h}{g}+\widetilde{\alpha}|$ and $2|\beta|$ remain unchanged, while $\sqrt{2}|\frac{h}{g}+\widetilde{\alpha}+\beta|$, $\sqrt{2}|\frac{h}{g}+\widetilde{\alpha}-\beta|$, $\sqrt{2}|\frac{h}{g}+\widetilde{\alpha}+\jmath\beta|$, $\sqrt{2}|\frac{h}{g}+\widetilde{\alpha}-\jmath\beta|$ are varied with the phase of $\beta$ (i.e., $\theta$). For convenience, we denote $\frac{h}{g}+\widetilde{\alpha}$ by $\xi$. According to the constraints $\mathrm{C1}$ and $\mathrm{C2}$, we have $\theta\in(-\pi,0)$. Specifically, we consider three cases, i.e., $-\frac{1}{4}\pi\leq\theta\leq0$, $-\frac{3}{4}\pi\leq\theta\leq-\frac{1}{4}\pi$, and $-\pi\leq\theta\leq-\frac{3}{4}\pi$. Due to space limit, we only discuss the first case, the other two cases can be proved in a similar way. When $-\frac{1}{4}\pi\leq\theta\leq0$, we have  
    	\begin{equation}
    	\sqrt{2}|\xi-\beta|=\sqrt{2}\min\left\{\lvert\xi+\beta\rvert,|\xi-\beta|,|\xi+\jmath\beta|,|\xi-\jmath\beta|\right\}. \nonumber
    	\end{equation}
     Then $D_{\min}$ is simplified as $D_{\min}\triangleq\min\left\{2|\xi|,2|\beta|,\sqrt{2}|\xi-\beta|\right\}$. We first compare the two distances $2|\beta|$ and $\sqrt{2}|\xi-\beta|$. By solving the equality $2|\beta|=\sqrt{2}|\xi-\beta|$ with $\beta = |\beta|e^{\jmath\theta}$, we can obtain the boundary between them, given by $|\beta_{1}|= |\xi|\left(-\cos(\theta)+\sqrt{1+\cos^{2}(\theta)}\right)$.
    	We further consider the following two subcases.
    	\begin{itemize}
    		\item \emph{Subcase 1}: $0\leq|\beta| \leq |\beta_{1}|$. In this case, we know that $2|\beta|\leq \sqrt{2}|\xi-\beta|$.
    	$D_{\min}$  is rewritten as $\min\left\{2|\xi|,2|\beta|\right\}$. Obviously, we have $D_{\min}=2|\beta|$ since $|\beta_{1}|\leq |\xi|$.
    		\item \emph{Subcase 2}: $|\beta|\geq|\beta_{1}|$. We have $2|\beta|\geq \sqrt{2}|\xi-\beta|$.
    		$D_{\min}$ is recast as $\min\left\{2|\xi|,\sqrt{2}|\xi-\beta|\right\}$. By solving the equality $2|\xi|=\sqrt{2}|\xi-\beta|$ with $\beta=|\beta|e^{\jmath\theta}$, we obtain another boundary of $\beta$, given by $|\beta_{2}| = |\xi|\left(\cos(\theta)+\sqrt{1+\cos^{2}(\theta)}\right)$.
    		Obviously, it follows that
    		\begin{itemize}
    			\item \emph{Subsubcase 1}: $|\beta_{1}|\leq|\beta|\leq|\beta_{2}|$, we have $D_{\min}=\sqrt{2}|\xi-\beta|$.
    			\item \emph{Subsubcase 2}: $|\beta|\geq|\beta_{2}|$, the minimum Euclidean distance is $D_{\min}=2|\xi|$.
    		\end{itemize}
    	\end{itemize}
Combining all the cases above together, Lemma \ref{lem-D-min-general} follows.
	\section{} \label{AppendexD}
		\begin{figure}[!t]
		\centering
		\captionsetup{font={scriptsize}}
		\subfigure[$|h/g|=0.8$, $\widetilde{\alpha}=0.3$] {
			\label{fig:general_beta_region_1}
			\includegraphics[width=0.2\textwidth]{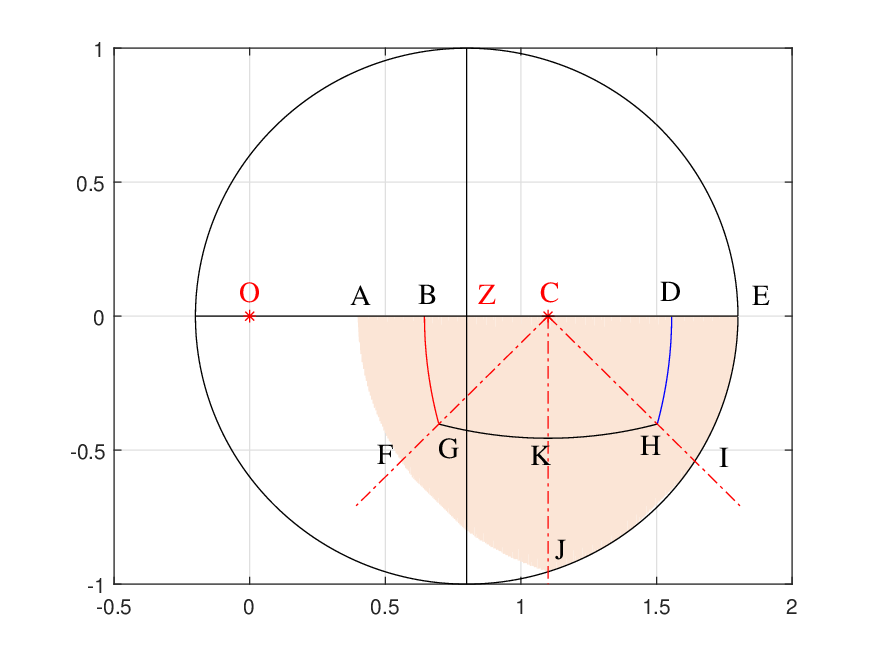}
		}
		\subfigure[$|h/g|=0.8$, $|\widetilde{\alpha}|=0.6$]{
			\label{fig:general_beta_region_2}
			\includegraphics[width=0.2\textwidth]{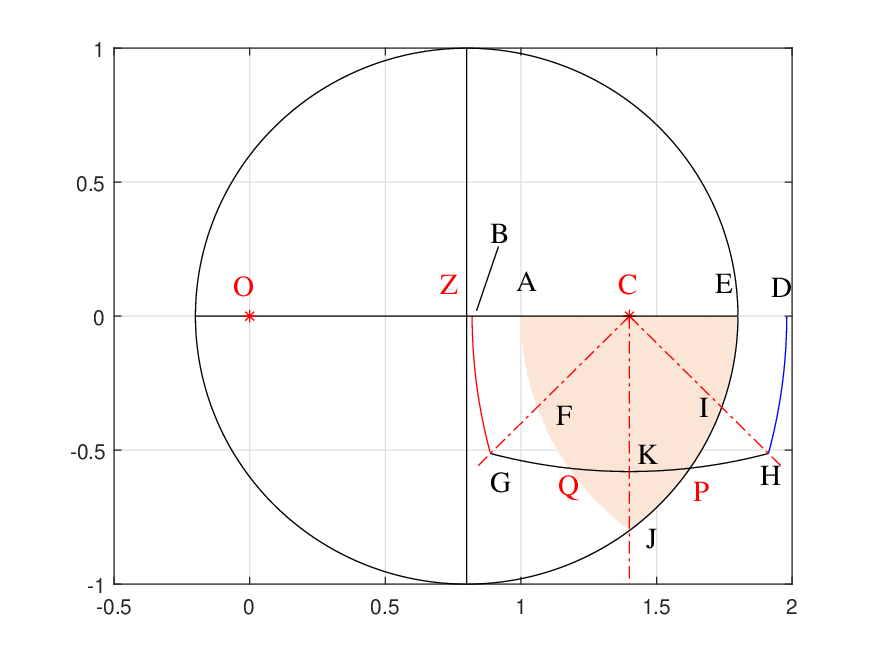}
		}
	\vspace{-0.3cm}
		\caption{ The feasible region of $\beta$ under $|h/g|=0.8$ and different $|\widetilde{\alpha}|$.}
	\end{figure}

     Here, we give an example for the case of $|h/g|=0.8$, as shown in Fig. \ref{fig:general_beta_region_1}, \ref{fig:general_beta_region_2}.  Based on $\mathrm{\overline{C2}}$ and $\mathrm{\overline{C4}}$, the feasible region of $\beta$ is given by the orange region. $\frac{h}{g}$ and $\widetilde{\alpha}$ are represented by the vector $O-Z$ and vector $Z-C$, respectively.
     Recall that in Lemma \ref{lem-D-min-general}, the region of $\beta$ is partitioned into five regions.
     From Fig. \ref{fig:general_beta_region_1}, $\mathbb{D}_{1}$, $\mathbb{D}_{2}$, $\mathbb{D}_{3}$, and $\mathbb{D}_{4}$ are represented by the regions closed by $B-D-H-G$, $D-E-I-H$, $F-G-H-I-J$, and $A-B-G-F$, respectively. $\mathbb{D}_{5}$ locates outside the orange region, and thus is not considered.
	 From Fig. \ref{fig:general_beta_region_1}-\ref{fig:general_beta_region_2}, there are two transition points for $\widetilde{\alpha}$, under which the boundaries of these small regions are changed. 
	 
	 1) The node $H$ coincides with $I$: This means $|CH|=|CI|$, yielding the first transition point $\widetilde{\alpha}_{1}$. It can be obtained by solving the equality $|\frac{h}{g}+\widetilde{\alpha}|\left(-\cos(\theta)+\sqrt{1+\cos^{2}(\theta)}\right)=\sqrt{1-|\widetilde{\alpha}|^2\sin^2(\theta)}-|\widetilde{\alpha}|\cos(\theta)$ under $\theta=-\frac{1}{4}\pi$, given by
	 \begin{small}
	 		 \begin{align}
	 	\widetilde{\alpha}_{1}&\approx0.25\left(\sqrt{8-0.54\left|h/g\right|^{2}}-1.27\left|h/g\right|\right).
	 	\end{align}
	 \end{small}	 
	 
	 2) The node $K$ coincides with $J$: This means $|CK|=|CJ|$, yielding the second transition point $\widetilde{\alpha}_{2}$. It can be obtained by solving the equality $|\frac{h}{g}+\widetilde{\alpha}|\left(\sin(\theta)+\sqrt{1+\sin^{2}(\theta)}\right)=\sqrt{1-|\widetilde{\alpha}|^2}$ under $\theta=-\frac{1}{2}\pi$, given by
	 \begin{small}
	 		\begin{align}
	 	\widetilde{\alpha}_{2}
	 	&\approx0.43\left(-0.34\left|h/g\right|+\sqrt{-0.68\left|h/g\right|^{2}+4.7}\right).
	 	\end{align}
	 \end{small}
	 
	 The above two transition points divide the $\widetilde{\alpha}$ into three intervals, i.e., $[0,\widetilde{\alpha}_{1})$, $[\widetilde{\alpha}_{1},\widetilde{\alpha}_{2}]$, $[\widetilde{\alpha}_{2},1]$.
	  Due to space limit, we only discuss the interval $[0,\widetilde{\alpha}_{1})$, seeing Fig. \ref{fig:general_beta_region_1} as an example.
	For subproblem \textbf{P3-1}, $2|\beta|$ is the minimum distance. Obviously, the optimal $\beta^{*}$ should be the vector $C-F$ or $C-I$ that maximizes the modulus of $\beta$. Due to the symmetry, we only consider $C-I$ for simplicity. 
	Similarly, for the subproblems \textbf{P3-2}, \textbf{P3-3} and \textbf{P3-4}, the optimal $\beta^{*}$ is given by the vector $C-H$, $C-G$, and $C-I$, respectively. Subproblem \textbf{P3-5} is not considered since the feasible region of $\beta$ is an empty set. By comparing the four optimal $\beta^{*}$ to the above four subproblems, we choose the one that maximizes the minimum Euclidean distance based on \eqref{eq:chooes-beta}, i.e., the vector $C-I$. Combing all cases together, Theorem \ref{Theorem-withD-beta} is thus proved.
    \section{} \label{AppendexE}
    Due to the symmetry of the $8$PSK constellation, the BER of signal $x$ can be obtained by calculating the error probability when $x_{1}$ is transmitted. According to \eqref{eq-ML-Detector-x}, we use $v=\frac{g^{H}y}{|g|^2}$ to make hard decision.
    As we know, $v$ can be rewritten as $v=x+\widetilde{z}$, where $\widetilde{z}\sim \mathcal{CN}(0,\widetilde{\sigma}^{2})$ and $\widetilde{\sigma}^{2}=\sigma^{2}/|g|^2$.
    Assuming that the $x_{1}$ locates at the point with coordinate $(1,0)$. 
    By introducing the polar coordinate $r = \sqrt{\Re\{v\}^{2}+\Im\{v\}^{2}}$, $\phi=\arctan \frac{\Im\{v\}}{\Re\{v\}}$, we can obtain the joint probability density function (PDF) of $r$ and $\phi$ by applying some transformations, given by $f(r,\phi)=\frac{r}{\pi\widetilde{\sigma}^{2}}e^{-\frac{r^{2}+1-2r\cos(\phi)}{\widetilde{\sigma}^{2}}}$.
    
    Then, the marginal PDF of $\phi$ can be obtained by taking integration over $v$, given by
    \begin{align}
    &f(\phi) = \int_{0}^{\infty}f(r,\phi)dv=\frac{r}{\pi\widetilde{\sigma}^{2}}e^{-\widetilde{\gamma_{b}}\sin^{2}(\phi)}\int_{0}^{\infty}e^{-\frac{(r-\cos(\phi))^2}{\widetilde{\sigma}^{2}}}dr \nonumber\\
    &\overset{a}{=}\frac{1}{2\pi}e^{-\widetilde{\gamma_{b}}\sin^{2}(\phi)}\int_{0}^{\infty}te^{-\frac{(t-\sqrt{2\widetilde{\gamma_{b}}}\cos(\phi))^2}{2}}dt \nonumber \\
    & \overset{b}{=} \frac{1}{2\pi}e^{-\widetilde{\gamma_{b}}}
    \!+\!\frac{1}{2}\sqrt{\frac{\widetilde{\gamma_{b}}}{\pi}}e^{-\widetilde{\gamma_{b}}\sin^{2}(\phi)}\cos(\phi) \left(1\!+\!\mathrm{erf}(\sqrt{\widetilde{\gamma_{b}}}\cos(\phi))\right) \nonumber
    \end{align}
    where $\widetilde{\gamma_{b}}=\frac{1}{\widetilde{\sigma}^{2}}=\frac{|g|^2}{\sigma^2}$, `$a$' follows by changing the integral variable $t=\frac{\sqrt{2}}{\widetilde{\sigma}}r$, `$b$' follows from the equality $\int_{0}^{\infty}ue^{-\frac{(u-u_{1})^{2}}{u_{2}}}du=\frac{1}{2}u_{2}e^{-\frac{u_{1}^{2}}{u_{2}}}+\frac{1}{2}u_{1}\sqrt{\pi u_{2}}\left(1+\mathrm{erf}\left(\frac{u_{1}}{\sqrt{u_{2}}}\right)\right)$.
    
    
    The probability of selecting $x_{5}$ as the decision  point when $x_{1}$ is transmitted, is given by
    \begin{align}
    P_{15}&=P(\hat{x}=x_{5} \big | x =x_{1}) = \int_{\frac{\pi}{8}}^{\frac{3\pi}{8}} f(\phi)d\phi  \nonumber\\
    &=\frac{1}{8}e^{-\widetilde{\gamma_{b}}}+\frac{1}{2}\sqrt{\frac{\widetilde{\gamma_{b}}}{\pi}}\int_{\frac{\pi}{8}}^{\frac{3\pi}{8}}\widetilde{f}(\phi)d\phi,
    \end{align}
    where $\widetilde{f}(\phi)=e^{-\widetilde{\gamma_{b}}\sin^{2}(\phi)}\cos(\phi) \left(1+\mathrm{erf}(\sqrt{\widetilde{\gamma_{b}}}\cos(\phi))\right)$. However, the integral term $\int_{\frac{\pi}{8}}^{\frac{3\pi}{8}}\widetilde{f}(\phi)d\phi$ is difficult to obtain its closed-form. For simulations, numerical integral is exploited to calculate this term.
    Similarly, we can obtain the probability $P_{12}$, $P_{16}$, and $P_{13}$, which are shown in \eqref{eq:P1} and \eqref{eq: P2}. Due to the symmetry, we have $P_{18}=P_{15}$, $P_{14}=P_{12}$, $P_{17}=P_{16}$. Combining all the probabilities above, the BER of signal $x$, $s$, and $c$ are obtained.
\end{appendices}
\bibliographystyle{IEEEtran}
\bibliography{IEEEabrv,refFile}


\end{document}